\documentclass[aps,pra,amssymb,twocolumn,showpacs,supersriptaddress,groupeaddress,floatfix]{revtex4-1}
\usepackage{amssymb}
\usepackage{amsmath}
\usepackage{graphicx}
\usepackage{float}
\usepackage{dcolumn}
\usepackage{bm}
\usepackage{natbib,hyperref}
\usepackage{hyperref}

\usepackage{color}
\hypersetup{colorlinks=true, 
    linkcolor=red,          
    citecolor=magenta,        
    filecolor=gree,      
    urlcolor=blue           
}

\usepackage[utf8]{inputenc}
\usepackage{graphicx}
\usepackage{dcolumn}
\usepackage{bm}
\usepackage{xcolor}
\usepackage{amsmath}
\usepackage{amssymb}
\usepackage{braket}
\usepackage{float}

\begin{document}

\title{Fast  quantum gates based on Landau-Zener-St\"uckelberg-Majorana transitions }


\author{Joan J. Cáceres$^{1}$}
\thanks{Current affiliation: Quantronics group, Université Paris-Saclay, CEA, CNRS, SPEC, 91191 Gif-sur-Yvette Cedex, France.}
\author{Daniel Domínguez$^{1}$ and María José Sánchez$^{1,2}$}
\affiliation{$^{1}$Centro At\'omico Bariloche and Instituto Balseiro (Universidad Nacional de Cuyo), 8400 San Carlos de Bariloche, R\'io Negro, Argentina.\\
$^{2}$Instituto de Nanociencia y Nanotecnolog\'{\i}a (INN),CONICET-CNEA, Argentina.}

\date{\today}

\begin{abstract}

Fast quantum gates   are of paramount importance for enabling efficient and error-resilient quantum computations. In the present work we analyze Landau-Zener-St\"uckelberg-Majorana (LSZM) strong driving protocols, tailored to implement fast gates with particular emphasis on small gap qubits. 
We derive  analytical equations to determine the  specific set of driving parameters 
for the implementation of single qubit and two qubit gates employing  single period sinusoidal pulses.
Our approach circumvents  the need to  scan experimentally a wide range of parameters and instead it allows to focus in fine-tuning the device near the analytically predicted values. We analyze the dependence of relaxation and decoherence on the 
 amplitude and frequency of the pulses, obtaining the optimal regime of driving parameters to mitigate the  effects of the environment.
 Our results focus on the study of the single qubit $X_{\frac{\pi}{2}}$,  $Y_{\frac{\pi}{2}}$ and identity gates. Also, we propose  the $\sqrt{\rm{bSWAP}}$  as the simplest two-qubit gate attainable through a robust LZSM driving protocol. 

\end{abstract}

\maketitle
\footnotetext{Afiliación actual para $^\dagger$: [Tu afiliación actual aquí]}
\section{Introduction}
One of the key ingredients in quantum computing is the ability to implement fast quantum gates, which are  the essential building blocks  to perform  quantum algorithms. In recent years, significant progress has been made in the design of quantum gates using superconducting qubits, which become one of the most promising platforms due to their scalability, long coherence times, and potential for fast high-fidelity operations\cite{krantz_2019,kjaergaard_2020,kwon_2021}.
The transmon qubit \cite{Koch_2007} and  the capacitively shunted flux qubit \cite{yan_2016}   establish   the basis for modern design of artificial atoms based  on superconducting circuits.  A more recent addition, the fluxonium,  \cite{manucharyan_2009,pop_2014,nguyen_2019,bao_2022, weiss_2022,somoroff_2023}, 
has a low transition frequency and a large anharmonicity, making it a promising candidate for quantum simulations and high-fidelity gate operations.

A variety of techniques for implementing fast quantum gates, including dynamical decoupling , composite pulses, and optimal control have been implemented so far \cite{leek_2007,bylander_2011,yang_2017,wang_2017,zhu_2021,shen_2021,ficheux_2021,bastrakova_2022,chen_2022}. 
These techniques have led to significant improvements in gate fidelity. However, most of them rely on  a resonant  Rabi driving with  frequency $\omega \approx \Delta=  E_1-E_0$ (the qubit energy gap)   and  small amplitude, $A\ll \Delta$, whose duration is adapted to perform the target operation (throughout this article we take $\hbar=1$).  
As  the gate time $t_g$  is inversely  proportional to the generalized Rabi frequency, i.e. $t_g \propto \frac{1}{\Omega_{R}} \propto \frac{1}{A}$  \cite{bloch_1940}, these schemes  usually  have limited gate speed involving time scales that are in conflict with those imposed by decoherence processes.

One approach to mitigating decoherence is to reduce the qubit coupling to the environment by using a low frequency, or small gaps qubit, as the heavy fluxonium. However, again in these cases, methods based on the Rabi resonant control  would be  unfeasible as  $ \Delta$ is small  and  in the rotating wave approximation (RWA)  as $A\ll \Delta$  results $t_g\gg t_\Delta=\frac{2\pi}{\Delta}$.

To circumvent the mentioned limitations,  alternatives beyond the resonant Rabi protocol have been recently proposed to experimentally implement fast gates with the decoherence time $t_{decoh}\gg t_{g}$\cite{avi_2014,campbell_2020,zhang_2021,petrescu_2023}. 
One of these  schemes  is  based on
 driving a composite qubit, formed from two capacitively coupled transmon qubits  which has a small gap  between two energy levels\cite{campbell_2020}. 
The qubit is controlled by a  Landau-Zener-St\"uckelberg-Majorana (LZSM) driving protocol, which  consist on driving the qubit with a strong amplitude and/or an off resonant harmonic signal \cite{Oliver_2005,Sillanpaa_2006,Ferron_2012,shevchenko_2012, ivakhnenko_2023}.
LZSM protocols  have been  successfully implemented in interferometry of superconducting qubits \cite{Oliver_2009, Berns_2008}, temporal oscillations \cite{bylander_2009} and  used in the quantum simulation of universal conductance fluctuations and weak localization phenomena \cite{gustavsson_2013, gramajo_2020}.

On the theoretical side, the study of the LZSM  driving protocols requires the implementation of  numerical methods, the  most useful  based on the Floquet formalism \cite{Shirley_1965,son_2009, Ferron_2012,Ferron_2016}, as the RWA is valid in the weak driving and resonant cases, but breaks down in the strong driving regime where analytical and perturbative approaches fail.  It is well known that the counter rotating   terms
lead to the shifts of resonances (Bloch-Siegert shift) and additional beat patterns in
the time evolution  \cite{bloch_1940,Shirley_1965} which are not captured in the  RWA.
In a quite recent paper \cite{yan_2015},  the Bloch-Siegert shift  was analytically  obtained along the entire driving-strength regime, i.e. for $0 < A/\omega < \infty $, by a simple 
analytical method based on an unitary transformation.
The method uses  a counter-rotating hybridized rotating wave approximation (CHRW) \cite{lu_2012,yan_2015,chen_2022}  and enables to obtain an effective  description of the qubit dynamics
which reproduces the numerical results
not only when the driving strength
is moderately weak but also for strong driving strengths, far beyond the perturbation theory.

In the present work, we  conduct an analysis of LZSM strong driving protocols suitable for implementing quantum gates in small gap qubits. By presenting precise analytical equations based on the CHRW approximation, we offer a method to determine the driving parameters (amplitude, frequency, initial and final idling times) required for both single qubit gates and the $\sqrt{bSWAP}$ gate. The approach eliminates the need for extensive experimental parameter scanning, allowing  to concentrate on fine-tuning the device based on the analytically predicted parameters. We suggest the  $\sqrt{bSWAP}$ gate as an ideal two-qubit gate achievable through a straightforward single one-period sinusoidal pulse using the strong driving LZSM protocol.

 The paper is organized as follows: In Sec.\ref{s2} we introduce  the counter-rotating hybridized rotating wave approximation (CHRW) to analyze the effective dynamics of  the driven qubit  Hamiltonian in terms of the operator $U(T)$, for a single period $T$ of a sinuosoidal drive. As a figure of  merit for the accuracy of the CHRW we  compute the error of this approximation, defined in terms of the Fidelity of the evolution with the operator $U(T)$ with respect to a target  (exact numerically computed)  unitary operator. 
 In Secs. \ref{sec:sqg} and \ref{sec:2qg} we analyze  the implementation of single qubit   $X_{\frac{\pi}{2}}$,  $Y_{\frac{\pi}{2}}$ and two qubit $\sqrt{bSWAP}$ gates respectively, with special focus on the determination of  the optimal  driving parameters in order  to engineering  fast gates with strong non resonant   LZSM protocols based on single period sinusoidal drives.
 The effect of relaxation and decoherence on the gate dynamics is analyzed in Sec.\ref{sec:dr}.
 Finally the summary and conclusions are presented in Sec.\ref{sec:concl}.






\section{Effective dynamics for strongly driven qubits}
\label{s2}


We start by considering the  standard two-level Hamiltonian modeling a driven qubit:
\begin{equation}
	H(t)=-\frac{\Delta}{2}\hat{\sigma}_{z}-\frac{\epsilon(t)}{2}\hat{\sigma}_{x},
 \label{eq:hamil}
\end{equation}
where $\hat{\sigma}_{z}$ and $\hat{\sigma}_{z}$ are the Pauli matrices and $\Delta$ is the qubit energy gap 
The Hamiltonian is written in the  basis spanned by $|0\rangle$ and
$|1\rangle$,  which correspond to the ground and excited states of the qubit respectively.
We consider a  transverse driving protocol $\epsilon(t)=A\sin(\omega t)$, as used 
recently in small gap qubits like superconducting composite qubits \cite{campbell_2020} and heavy fluxonium qubits \cite{weiss_2022,zhang_2021}. 

For large gap qubits, like the transmon, the driving strength $A$ is typically small compared with the qubit gap $\Delta$, and qubit control and quantum gates are implemented in the Rabi-driving regime at resonant frequencies $\omega\approx\Delta$, that can be accurately described within the RWA.  The recent development of highly coherent  qubits with small gaps requires  operation with non resonant  fast  drives $\omega>\Delta$  and large driving amplitudes, using control protocols based on LZSM transitions \cite{Oliver_2005,Oliver_2009,Ferron_2012,shevchenko_2012,ivakhnenko_2023}.
In this case, and in contrast with the Rabi-driving protocol, there are no simple expressions to predict the driving parameters needed for the implementation of quantum gates. Instead, calibration protocols scanning the pulse amplitude $A$ and  frequency $\omega$ are usually performed in the experiments \cite{campbell_2020,weiss_2022,zhang_2021}. 
For instance in Ref.\cite{campbell_2020}, as a first step in the calibration protocol, 
the transition probability from the ground state to the excited state  after a single period  $T=2\pi/\omega$ of the sinusoidal drive, $P_{01}\equiv P_{|0\rangle\rightarrow|1\rangle}(T)$, is measured as a function of $A$ and $\omega$. From this scanned transition probability, values of $A$ and $\omega$ are chosen such that  for these values $P_{01}$   corresponds to the implementation  of a given quantum gate. In a second calibration step, idling times before and after the driving pulse are finely tuned-up in order to implement the desired quantum gate \cite{campbell_2020}.

\begin{figure}
    \centering
    \includegraphics[width=\linewidth]{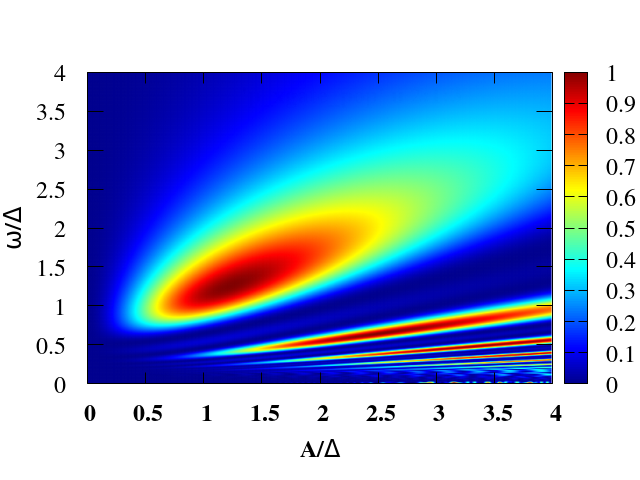}
    \caption{Transition probability $P_{01}=|\langle 1| U_{num}(T)|0\rangle|^2$ computed numerically using a fourth order Trotter-Suzuki algorithm. Intensity plot of $P_{01}$ as a function of the driving frequency $\omega$ and amplitude $A$, both normalized by the qubit gap $\Delta$. See text for details.}
    \label{fig:fig1}
\end{figure}

We have computed  the time evolution with the $H(t)$ of Eq.(\ref{eq:hamil}) using a fourth order Trotter-Suzuki algorithm \cite{hatano2005}, to compare with the calibration protocol of Ref.\cite{campbell_2020}. From the time evolution operator computed numerically, $U_{num}(t)$, we obtain $P_{01}=|\langle 1| U_{num}(T)|0\rangle|^2$ and plot it  in Fig.\ref{fig:fig1} as a function of $A$ and $\omega$, scanning the same range of values as in the experiment.  The computed probabilities align closely with the experimental results shown in  Fig.S3 of Ref.\cite{campbell_2020}.

An analytical accurate estimation of the amplitude and frequency of the driving pulse could greatly simplify the calibration procedure.
The difficulty is that  large driving strengths require to go beyond the RWA  and to account for counter-rotating effects.
To obtain an effective  description of the qubit dynamics  an strategy is  to use  the  CHRW approximation \cite{lu_2012,yan_2015,chen_2022}, applied to this case. To this end, we start with the transformation $|\psi'(t)\rangle=U_x|\psi(t)\rangle$,
 $H'=U_xHU_x^\dagger+i(\partial_t U_x)U_x^\dagger$, with $U_{x}=e^{-i\frac{\phi}{2}\hat{\sigma}_{x}}$. This gives \cite{yan_2015}
\begin{equation}
	H'=-\frac{(\epsilon-\dot{\phi})}{2}\hat{\sigma}_{x}-\frac{\Delta}{2}(\cos\phi \hat{\sigma}_{z}-\sin\phi \hat{\sigma_{y}}).
 \label{eh1}
\end{equation}
We take $\dot{\phi}=\xi\epsilon$, with $\xi$ a parameter to be  determined later.  Thus $\phi(t)=-\xi\frac{A}{\omega}\cos(\omega t)$ and  $\exp{({i\phi})}=\exp{(-i\xi\frac{A}{\omega}\cos(\omega t))}$.
Using the Jacobi-Anger expansion in terms of Bessel functions for:
\begin{equation*}
e^{ix\cos a}=\sum_{k=-\infty}^{k=+\infty}i^{k}J_{k}(x)e^{ika}=J_{0}(x)+\sum_{k=1}^{\infty}2i^{k}J_{k}(x)\cos(k a),
\end{equation*}
we approximate in Eq.(\ref{eh1}) (to lowest order in the Fourier expansion):
\begin{eqnarray}
\cos\phi&\approx& J_{0}(\xi\frac{A}{\omega})\nonumber\\
\sin\phi&\approx&-2J_{1}(\xi\frac{A}{\omega})\cos(\omega t),
\end{eqnarray}
and therefore
\begin{equation*}
H'=-\frac{\epsilon(1-\xi)}{2}\hat{\sigma}_{x}-\frac{\Delta}{2}\left[J_{0}(\xi\frac{A}{\omega})\hat{\sigma}_{z}+2J_{1}(\xi\frac{A}{\omega})\cos(\omega t)\hat{\sigma_{y}}\right],
\end{equation*}
which can be rewritten as:
\begin{equation}
H'=-\frac{\widetilde{\Delta}}{2}\hat{\sigma}_{z}-\frac{1}{2}\left[A(1-\xi)\sin(\omega t)\hat{\sigma}_{x}+2 \Delta J_{1}(a)\cos(\omega t)\hat{\sigma_{y}}\right],
\label{eqhprime}
\end{equation}
with $\widetilde{\Delta}=\Delta J_{0}(a)$, and $a=\xi A/ \omega$.

After expressing $\xi$ in terms of the self-consistent equation:
\begin{equation}
A(1-\xi)=2\Delta J_{1}(\xi\frac{A}{\omega})=\widetilde{A},
\label{eqconsist}
\end{equation}
we can rewrite Eq.(\ref{eqhprime}) as:
\begin{eqnarray}
H'=-\frac{\widetilde{\Delta}}{2}\hat{\sigma}_{z}-\frac{\widetilde{A}}{2}\left[\sin(\omega t)\hat{\sigma}_{x}+\cos(\omega t)\hat{\sigma_{y}}\right]\\ \nonumber
=-\frac{\widetilde{\Delta}}{2}\hat{\sigma}_{z}-\frac{\widetilde{A}}{2}\left[-i e^{i\omega t}\hat{\sigma}_{+}+i e^{-i\omega t}\hat{\sigma_{-}}\right].
\end{eqnarray}
which  can be solved exactly. 
After transforming with the unitary operator $R=e^{-i\frac{\omega t}{2}\hat{\sigma}_{z}}$, we obtain:

\begin{eqnarray}
H''=RH'R^{\dagger}+i\frac{d R}{dt}  R^{\dagger} \nonumber\\
=-\frac{\widetilde{\delta}}{2}\hat{\sigma}_{z}-\frac{\widetilde{A}}{2}\hat{\sigma_{y}},
\label{eqnhdtilde}
\end{eqnarray}
being $\widetilde{\delta}=\widetilde{\Delta}-\omega$. 

Equation (\ref{eqnhdtilde}) can be easily  diagonalized with the transformation $W=e^{-i\frac{\theta}{2}\hat{\sigma}_{x}}$, being
$\tan\theta=\widetilde{A}/\widetilde{\delta}$, obtaining:
\begin{equation}
H_{d}=WH''W^{\dagger}=-\frac{\Omega_{R}}{2}\hat{\sigma}_{z},
\end{equation}
with 
$$\Omega_{R}=\sqrt{\widetilde{\delta}{{}^2}+\widetilde{A}{{}^2}}=\sqrt{[\Delta J_{0}(\xi\frac{A}{\omega})-\omega]^{2}+4\Delta^{2}J_{1}{{}^2}(\xi\frac{A}{\omega})},$$
the generalized Rabi frequency. 

Taking into account the previous transformations,  the  evolution operator associated to Eq.(\ref{eq:hamil}), in the CHRW approximation, results :
\begin{equation}
U(t)=U_{x}^{\dagger}(t)R^{\dagger}(t)W^{\dagger}e^{i\frac{\Omega_{R}t}{2}\hat{\sigma}_{z}}WR(0)U_{x}(0).
\end{equation}

In our case, for the implementation of fast quantum gates, we are interested in the evolution after one period of the driving, $T$,  
which is:
\begin{equation}
U(T)=-e^{i\frac{\theta-a}{2}\hat{\sigma}_{x}}e^{i\frac{\pi\Omega_{R}}{\omega}\hat{\sigma}_{z}}e^{-i\frac{\theta-a}{2}\hat{\sigma}_{x}}
\label{eqUT1}
\end{equation}
\begin{equation*}
=-\left(\begin{array}{cc}
	\cos\alpha+i\sin\alpha\cos{\tilde\theta} & \sin\alpha\sin{\tilde\theta}\\
	-\sin\alpha\sin{\tilde\theta} & \cos\alpha-i\sin\alpha\cos{\tilde\theta}
\end{array}\right)\nonumber,
\end{equation*}
with $\alpha=\frac{\pi\Omega_{R}}{\omega}$ and  ${\tilde\theta}=\theta-a=\theta-\xi\frac{A}{\omega}$.
The transition probability between the qubit states, $\left|0\right\rangle \rightarrow\left|1\right\rangle $,
can then be obtained in this approximation as,

\begin{eqnarray}
P_{01}&=&|\langle 1| U(T)|0\rangle|{{}^2}=\sin{{}^2}{\tilde\theta}\sin{{}^2}\alpha
\label{eqP01}\\
&=&\frac{\left[\widetilde{A}\cos(\xi\frac{A}{\omega})-\widetilde{\delta}\sin(\xi\frac{A}{\omega})\right]{{}^2}}{\Omega_{R}^{2}}\sin{{}^2}\frac{\pi\Omega_{R}}{\omega}.\nonumber
\end{eqnarray}

\begin{figure}[ht]
    \centering
    \includegraphics[width=\linewidth]{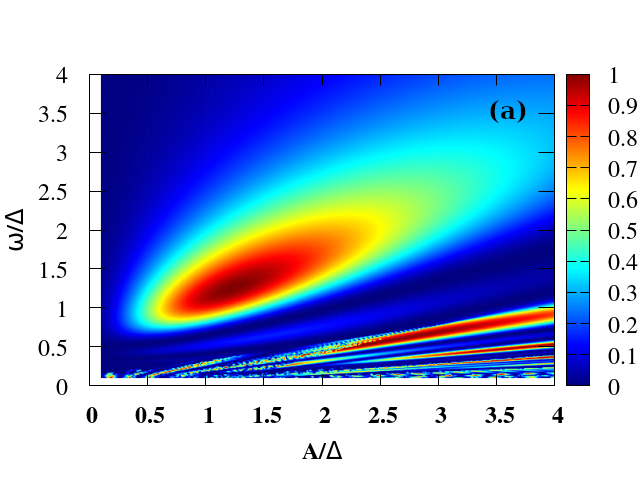}
    \includegraphics[width=\linewidth]{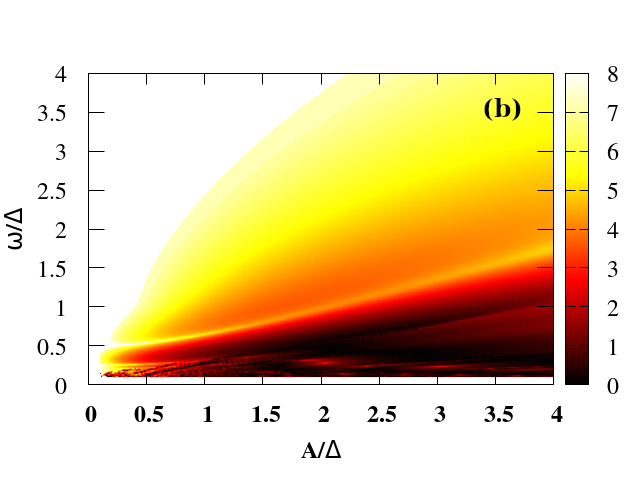}
    \caption{(a) Transition probability $P_{01}$ obtained from Eq.(\ref{eqP01}) as a function of  driving frequency $\omega/\Delta$ and amplitude $A/\Delta$. (b) Plot of the error  ${\cal E}$ of the CHRW approximation.  The color scale of the intensity plot corresponds to $-\log_{10}|{\cal E}|$. Regions in white color correspond to ${\cal E}<10^{-8}$.}
    \label{fig:fig2}
\end{figure}

In Fig.\ref{fig:fig2}(a)  we plot the transition probability computed from the analytical expression Eq.(\ref{eqP01})  as a function of the amplitude of the driving $A/\Delta$ and the frequency $\omega/\Delta$, (both normalized in terms of the qubit gap).  The agreement with the numerical result of Fig.(\ref{fig:fig1})   is remarkable, despite some differences noticeable in the range of small  $\omega/A  $,  due to numerical instabilities in the solution of  Eq.(\ref{eqconsist})  originated in the highly oscillatory behaviour of the   Bessel function   $J_1(x)$ for large values of  its argument $x$.


In order to check the accuracy of the CHRW approximation we compare the approximated $U(T)$  of Eq.(\ref{eqUT1}) with the  
the numerically exact $U_{num}(T)$. We  quantify  the error of the approximation  as 
\begin{equation}
    {\cal E}= 1- {\cal F} = 1- \frac{{\rm Tr}(U^\dagger U)+|{\rm Tr}(U_{tg}^\dagger U)|^2}{d(d+1)}\,. 
    \label{eq:fidel}
\end{equation}
where ${\cal F}$ is the standard expression for the fidelity of an evolution operator $U$ with  respect to a target unitary operator $U_{tg}$ \cite{pedersen_2007}.
In the present  case is $U_{tg}=U_{num}(T)$,  with $d=2$ the dimension of the space, 
and  ${\rm Tr}(U^\dagger U)=d$, as $U(T)$ is exactly unitary.
In Fig.\ref{fig:fig2}(b)  we plot the error ${\cal E}$. As expected  \cite{lu_2012}, the CHRW approximation is very accurate 
in the range $\omega \gtrsim A/2$ and $\omega \gtrsim \Delta$ (with  ${\cal E}\ll 10^{-3}$), which  is also   the range of interest for the experiment of Ref.\cite{campbell_2020}. Other methods of approximation as the Magnus expansion (used  in Ref.\cite{weiss_2022}) and the RWA in a double rotating frame (used in  Ref.\cite{deng_2015,deng_2016}) are much less accurate (see Appendix B for a comparison of   the different approximations).



In the following sections we shall analyse different implementations of single and two qubit gates following this  driving protocol.


\section{Implementation of single qubit gates }\label{sec:sqg}

Here we analyze the conditions to implement fast single qubit gates with a strong driving  protocol based on non resonant sinusoidal pulses 
(see Eq.(\ref{eq:hamil})).

First, we note that $Z_\alpha=\exp(-i\alpha \hat{\sigma}_{z}/2) $ gates, can be realized by ``idling" operations in the time  evolution with the qubit set at $\epsilon=0$ for a time $t=\alpha/\Delta$, as implemented in Ref.\cite{campbell_2020}.
In addition to the continuous varying $Z_\alpha$ gate a complete set of single qubit gates can be realized,  implementing  for instance  $X_{\pm\frac{\pi}{2}}= \exp(\mp i\pi\hat{\sigma}_{x}/4)$ and $Y_{\pm\frac{\pi}{2}}= \exp(\mp i\pi \hat{\sigma}_{y}/4)$ gates. 

In the case of the   $Y_{\frac{\pi}{2}}$ gate, we  can write it  in matrix form as:
\begin{equation}
Y_{\frac{\pi}{2}}=\frac{\sqrt{2}}{2}\left(\begin{array}{cc}
	1 & -1\\
	1 & 1
\end{array}\right).
\label{eq:ygate}
\end{equation}
A direct comparison of  Eq.(\ref{eq:ygate}) with the CHRW expression for   the operator $U(T)$, Eq.(\ref{eqUT1}), gives the following condition to  implement the $Y_{\frac{\pi}{2}}$ gate with the largest $\omega$:
\begin{eqnarray}
\label{parygate}
\omega&=&\frac{4}{3}\Omega_{R},
\\
\theta-\xi\frac{A}{\omega}&=&\frac{\pi}{2}
\nonumber
\end{eqnarray}
Numerical solution of these equations together with Eq.(\ref{eqconsist}) give $\omega_Y\approx 2.07\Delta$ and $A_Y\approx 2.87\Delta$. The general conditions  for the $Y_{\frac{\pi}{2}}$ gate are: $\omega=\Omega_R/(2k+3/4),\; {\tilde\theta}=(2l+1/2)\pi$ and 
$\omega=\Omega_R/(2k+5/4),\; {\tilde\theta}=(2l+3/2)\pi$, for $k,l$ integers. Solutions with $k\not=0, l\not=0$ give low $\omega$ and large $A$, beyond the parameter range for the CHRW approximation.

\begin{figure}[th]
	\centering
	\includegraphics[width=\linewidth]{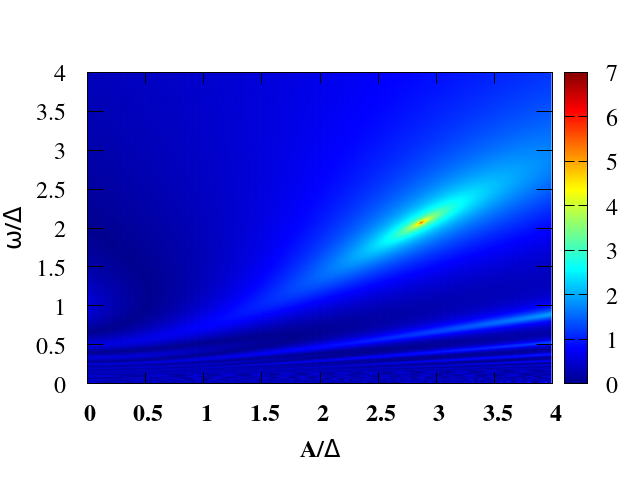}
	\caption{Location of the parameters for implementation of the $Y_{\frac{\pi}{2}}$ gate.
 Plot of the error function ${\cal E}$ that compares the numerically exact evolution operator $U_{num}$ with $Y_{\frac{\pi}{2}}$.  The color scale of the intensity plot corresponds to $-\log_{10}|{\cal E}|$. The point where ${\cal E}<10^{-7}$ (practically zero within numerical precision) gives the operational parameters  of the  gate,  frequency $\omega_Y$ and amplitude $A_Y$. It agrees with the analytical estimate of Eq.(\ref{parygate}), $\omega_Y\approx 2.07\Delta$ and $A_Y\approx 2.87\Delta$. }
     \label{fig:fidel_ypi2}
\end{figure}

It  is clear that the analytical estimate of the operational parameters for the gate implementation avoids the experimental cost of scanning parameters in a wide range. 
In the experiment, as described in the previous section, the fast gate is implemented applying the sinusoidal pulse 
for a single period $T=2\pi/\omega$, and  the corresponding values of $A$ and $\omega$ are  determined by performing different measurements  scanning the amplitude and frequency. In a similar way, we can compute numerically the exact evolution operator $U_{num}(T)$ varying the parameters $\omega$  and $A$,  for instance in the range $[0,4\Delta]$. To obtain the numerically exact conditions for the $Y_{\frac{\pi}{2}}$ gate, we show in Fig.\ref{fig:fidel_ypi2} the  error function ${\cal E}$ from Eq.(\ref{eq:fidel}) with $U=U_{num}(T)$ compared with the target $U_{tg}=Y_{\frac{\pi}{2}}$. The point of minimum  ${\cal E}$  (which is  near the numerical precision for our calculation of $U_{num}(T)$, ${\cal E}\approx 10^{-7}$)
 corresponds to the  operational point in $A,\omega$ for implementing the gate $Y_{\frac{\pi}{2}}$. This point   agrees very accurately with the values $(A_Y, \omega_Y)$ computed previously from Eq.(\ref{parygate}).

In the case of the $X_{\frac{\pi}{2}}$ gate, its matrix representation is
\begin{equation}
\label{eq:xgate}
X_{\frac{\pi}{2}}=\frac{\sqrt{2}}{2}\left(\begin{array}{cc}
	1 & -i\\
	-i & 1
\end{array}\right).
\end{equation}
A comparison with the  approximate $U(T)$ given in  Eq.(\ref{eqUT1}) shows that  $X_{\frac{\pi}{2}}$  can not be realized directly,  since  $\sin\alpha\sin{\tilde\theta}\neq i$.  However, one can add after the sinusoidal pulse an idle  (Z gate) evolution \cite{mckay_2017,campbell_2020} during a  time $t_{f}=\frac{\pi}{\Delta}$, such that  $e^{i\frac{\Delta}{2}\hat{\sigma}_{z}t_f}=i\hat{\sigma}_{z}$.
Then, after the complete evolution given by  $U(T+t_{f})=i\hat{\sigma}_{z}U(T)$,  the main conditions to implement a  $X_{\frac{\pi}{2}}$  gate result:
\begin{eqnarray}
\label{parxgate}
\omega&=&2\Omega_{R},\\
\theta-\xi\frac{A}{\omega}&=&\frac{3\pi}{4}.\nonumber
\end{eqnarray}
Numerical solution of these equations together with Eq.(\ref{eqconsist}) give $\omega_X\approx 0.81\Delta$ and $A_X\approx 0.68\Delta$. The general conditions  the for the $X_{\frac{\pi}{2}}$ gate are: $\omega=\Omega_R/(2k+1/2),\; {\tilde\theta}=(2l+3/4)\pi$ and 
$\omega=\Omega_R/(2k+3/2),\; {\tilde\theta}=(2l+5/4)\pi$, for $k,l$ integers.

\begin{figure}
    \centering
    \includegraphics[width=\linewidth]{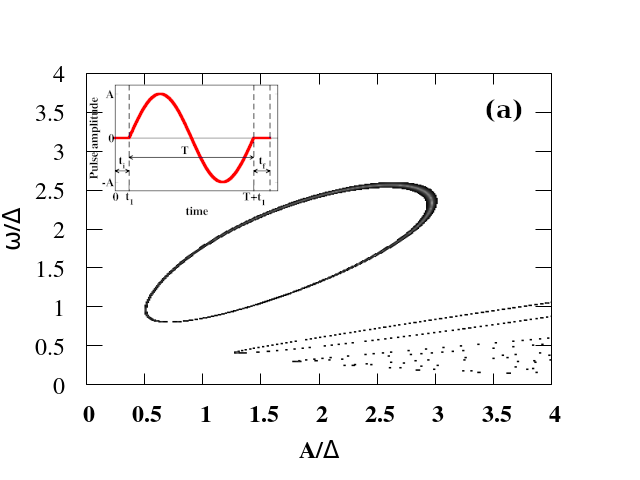}    
    \includegraphics[width=\linewidth]{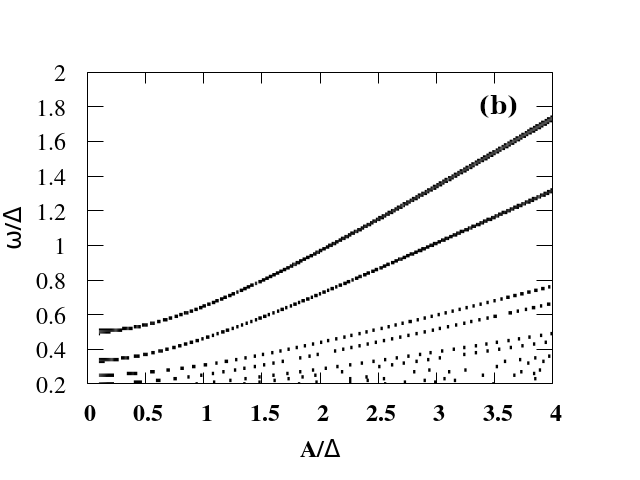}
    \caption{(a) Plot of the parameter sets  $\omega$  and $A$ for the implementation of $X_{\frac{\pi}{2}}$ and $Y_{\frac{\pi}{2}}$ gates, obtained from the analytical expressions of Eqs.(\ref{eqPXY}) and (\ref{eq:xiapp}) (except for the low frequency points, that correspond to the numerical evaluation of the condition $P_{01}=1/2$). See text for details.
     Inset: Schematic representation of the pulsing protocol with an idle time $t_i$ added before the sinusoidal drive and a second idle time $t_f$ afterwards. (b) Values of $\omega$  and $A$ to implement the identity operation
   determined from  Eq.(\ref{eqIdent}). }
    \label{fig:figsin}
\end{figure}

The problem with the above conditions, Eqs.(\ref{parygate}) and (\ref{parxgate}), is that each one requires  very specific (and {\it different}) frequencies ($\omega_Y$  and $\omega_X$)
and  gate times ($T_Y=2\pi/\omega_Y$ and $T_X=2\pi/\omega_X$) to implement  them.

A more general procedure \cite{mckay_2017,campbell_2020}, that expands the possibilities in parameter space, is to add an idle time $t_{i}$ before the sinusoidal drive and a second idle time $t_{f}$ afterwards, see inset in  Fig.\ref{fig:figsin}(a). Calling 
$\tau_{+}=(t_{i}+t_{f})\Delta/2$ 
and  $\tau_{-}=(t_{f}-t_{i})\Delta/2$, 
the evolution operator in the CHRW approximation results
\begin{equation}
U(t_{i}+T+t_{f})=
\nonumber
\end{equation}
\begin{equation*}
	\! \! \! \! \! =\left( \! \begin{array}{cc}
		-e^{i\tau_{+}}(\cos\alpha+i\sin\alpha\cos{\tilde\theta}) & -e^{i\tau_{-}}\sin\alpha\sin{\tilde\theta}\\
		e^{-i\tau_{-}}\sin\alpha\sin{\tilde\theta} & -e^{-i\tau_{+}}(\cos\alpha-i\sin\alpha\cos{\tilde\theta} )
	\end{array}\! \! \! \right) \! . \nonumber
\end{equation*}
For both gates, $X_{\frac{\pi}{2}}$ and $Y_{\frac{\pi}{2}}$,  the transition probability after one period is $P_{01}=|\langle 1|U|0\rangle|^2=1/2$, which corresponds to   the implicit  condition for  $\omega$ and $A$ given from the equation,
\begin{equation}
   \sin{{}^2}{\tilde\theta}\sin{{}^2}\alpha = \frac{1}{2}.
\label{eqPXY}
\end{equation}
Thus after imposing this condition  in $U(t_{i}+T+t_{f})$  one gets  
\begin{equation}
U (t_{i}+T+t_{f})=\frac{\sqrt{2}}{2}\left(\begin{array}{cc}
	-e^{i(\tau_{+}+\nu)} & -e^{i\tau_{-}}\\
	e^{-i\tau_{-}} & -e^{-i(\tau_{+}+\nu)}
\end{array}\right)
\end{equation}
where 
\begin{equation}
    \tan\nu=\tan\alpha\cos{\tilde\theta}.
\label{eq:nu}
\end{equation}
Then   the $X_{\frac{\pi}{2}}$ gate can be obtained for 
\begin{eqnarray}
    \tau_{-}&=&(2n + 1/2)\pi\nonumber\\
    \tau_{+}&=&(2k+1)\pi-\nu,
\label{eq:xtaus}
\end{eqnarray}
 being $k$ and $n$  integers, while the $Y_{\frac{\pi}{2}}$ gate can be implemented for
\begin{eqnarray}
    \tau_{-}&=&2n\pi\nonumber\\
    \tau_{+}&=&(2k+1)\pi-\nu,
    \label{eq:ytaus}
\end{eqnarray}
after straightforward comparisons with Eqs.(\ref{eq:xgate}) and (\ref{eq:ygate}), respectively.

To summarize, in order to determine the driving parameters $\omega,A,t_i,t_f$ for the gate implementation one can proceed as follows. For a given driving frequency $\omega$ one determines the possible  amplitudes $A$  solving Eq.(\ref{eqPXY}). 
  Notice that since the relevant solutions of Eq.(\ref{eqPXY}) are for  $\omega > A$, it is very accurate to use for the $\xi$ parameter the expression
\begin{equation}
    \xi\approx \frac{\omega}{\omega+\Delta} ,
    \label{eq:xiapp}
\end{equation}  
 from a first order approximation of Eq.(\ref{eqconsist}).
The resulting curve in $A,\omega$ space is shown in Fig.\ref{fig:figsin}(a), where all the possible values for implementation of $X_{\frac{\pi}{2}}$ and $Y_{\frac{\pi}{2}}$ gates are plotted. We find that they fall  within the range  
$\Delta\lesssim\omega<3\Delta$ and $\Delta/2 \lesssim A < 3\Delta$.
(We also  plot for completeness in Fig.\ref{fig:figsin}(a) the low frequency curves, for $\omega \ll  A$,
even when  these cases are not of interest for the implementation of fast gates. These points were obtained from the evaluation of $P_{01}=|\langle 1|U_{num}|0\rangle|^2=1/2$ using the numerically exact evolution, since the CHRW approximation does not apply in this case.)

Once the chosen driving parameters $A, \omega$ are determined from Eq.(\ref{eqPXY}), the values of the idling times $t_i$ and $t_f$ needed to implement a $X_{\frac{\pi}{2}}$ or a $Y_{\frac{\pi}{2}}$ gate   can be obtained from Eqs.(\ref{eq:xtaus}) or (\ref{eq:ytaus}), respectively.
 
 It has been argued in Ref.\cite{weiss_2022} that, since different physical qubits could have different $\Delta$ parameters, it is useful to have {\it variable-time} single qubit identity operations, to be  able to perform operations in one qubit  avoiding that a second qubit acquires a dynamical phase at the same time. Comparing the CHRW expression of $U(T)$ given in Eq.(\ref{eqUT1}) with the identity matrix we obtain that the identity operation can be implemented for the $A,\omega$ that satisfy the simple condition 
 \begin{equation}
     \omega=\Omega_R/(2k+1) . \label{eqIdent}
 \end{equation}
 The resulting values of $A,\omega$ are shown in Fig.\ref{fig:figsin}(b), where we observe that in this case it is possible to use arbitrary large values of $\omega$ (and large $A$). On the other hand,  for the $X_{\frac{\pi}{2}}$ and $Y_{\frac{\pi}{2}}$ gates one can see in Fig.\ref{fig:figsin}(a) that there is an upper limit in the frequency range for their implementation.

\section{Two qubit gates} \label{sec:2qg}

Any universal quantum instruction set requires  the implementation of at least one entangling two-qubit gate \cite{huang_2023,krantz_2019,kwon_2021}.
Here we consider the parametrically driven two qubit Hamiltonian:
\begin{equation}
H_{2q}(t)=-\frac{\Delta_{1}}{2}\hat{\sigma}_{z}\otimes I-\frac{\Delta_{2}}{2}I\otimes\hat{\sigma}_{z}-\frac{\epsilon(t)}{2}\hat{\sigma}_{x}\otimes\hat{\sigma}_{x},
\label{eq:h2q}
\end{equation}
with driving in the coupling parameter $\epsilon(t)=A\sin(\omega t)$. 
Its matrix representation, using the  basis $\left\{ \left|00\right\rangle ,\left|01\right\rangle ,\left|10\right\rangle ,\left|11\right\rangle \right\} $
is 
\begin{equation}
H_{2q}(t)=-\frac{1}{2}\left(\begin{array}{cccc}
	\Delta_{1}+\Delta_{2} & 0 & 0 & \epsilon(t)\\
	0 & \Delta_{1}-\Delta_{2} & \epsilon(t) & 0\\
	0 & \epsilon(t) & \Delta_{2}-\Delta_{1} & 0\\
	\epsilon(t) & 0 & 0 & -\Delta_{1}-\Delta_{2}
\end{array}\right).\nonumber
\end{equation}
This two-qubit Hamiltonian with a $\hat{\sigma}_{x}\otimes\hat{\sigma}_{x}$ tunable coupling  has been implemented for example  in  coupled fluxonium qubits \cite{weiss_2022,moska2021,moska_2022}. 

In the following  we show that with a single period sinusoidal drive, it is straightforward to get the $\sqrt{\rm{bSWAP}}$ entangling gate \cite{poletto_2012,roth2017,nesterov2021}:

\begin{equation}
\label{eq:uent}
U_{ent}=\sqrt{\rm{bSWAP}}=\left(\begin{array}{cccc}
	\frac{\sqrt{2}}{2} & 0 & 0 & -\frac{\sqrt{2}}{2}\\
	0 & 1 & 0 & 0\\
	0 & 0 & 1 & 0\\
	\frac{\sqrt{2}}{2} & 0 & 0 & \frac{\sqrt{2}}{2}
\end{array}\right),
\end{equation}
which  generates the entangled states $(|00\rangle\pm|11\rangle)/\sqrt{2}$ and leaves invariant the subspace spanned by $\{|01\rangle,|10\rangle\}$. It is easy to show that it is locally equivalent to the $\sqrt{\rm{iSWAP}}$ gate \cite{poletto_2012,huang_2023}. 

\begin{figure}[p]
	\centering
    \includegraphics[width=\linewidth]{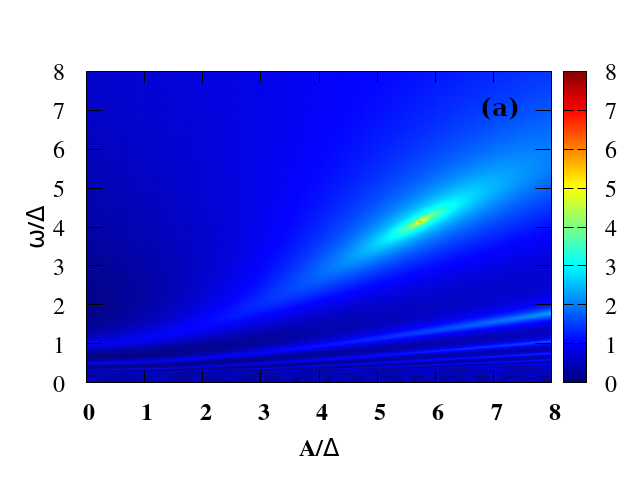}
	\includegraphics[width=\linewidth]{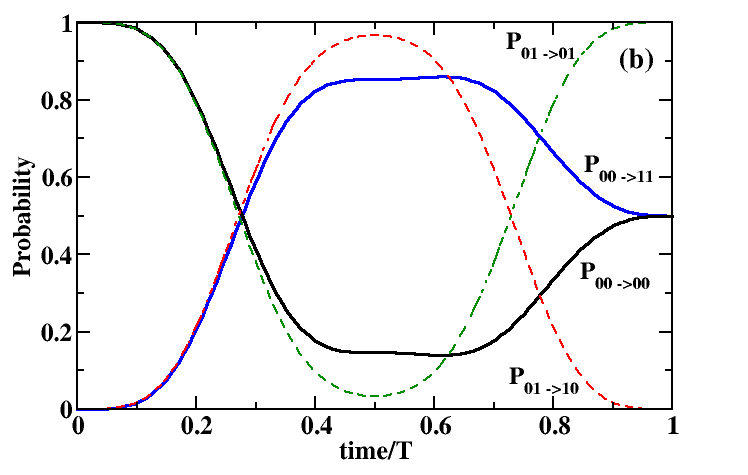}    
    \includegraphics[width=\linewidth]{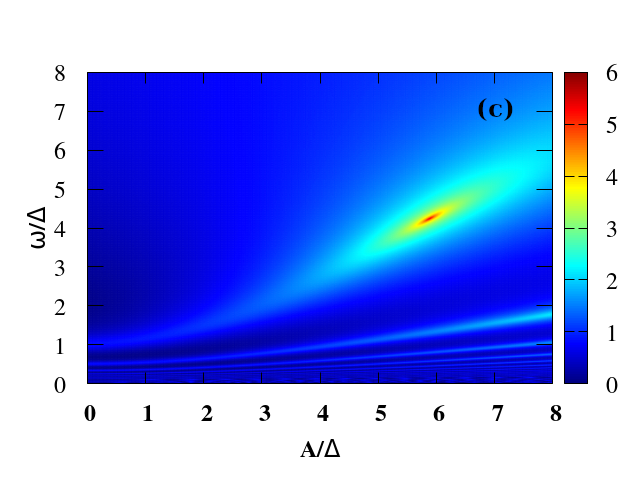}
	\caption{(a) 
 Plot, as  a  function of the parameters $A$ and $\omega$, of the error  ${\cal E}$  that  compares $\sqrt{\rm{bSWAP}}$ with the numerically exact $U_{num, 2q}$, computed  from $H_{2q}$ Eq.(\ref{eq:h2q}).  The color scale of the intensity plot corresponds to $-\log_{10}|{\cal E}|$. The point where ${\cal E}<10^{-7}$ (practically zero within numerical precision) gives the operational parameters  of the  gate,  frequency $\omega_{bS}$ and amplitude $A_{bS}$. It agrees with the analytical estimate of 
  $\omega_{bS}\approx4.14\Delta$, $A_{bS}\approx5.74\Delta$. (b) Time evolution  of the population transfers during a driving period for  $\omega_{bS},A_{bS}$. Continuous lines correspond to the evolution of the populations of the $|00\rangle$ (black) and the $|11\rangle$ states (blue), after the initial state $|00\rangle$. Dashed lines correspond to the evolution of the populations of the $|01\rangle$ (green) and the $|10\rangle$ states (red), after the initial state $|01\rangle$. (c)  Same as (a) but for two slightly different qubits with $\Delta_2=1.05\Delta_1$. See text for more details.}
 \label{fig:fidel2q}
\end{figure}

The $\sqrt{\rm{bSWAP}}$ gate can be exactly implemented in the ideal case when both qubits have equal gaps, $\Delta_{1}=\Delta_{2}=\Delta$. To demonstrate its realization,  we calculate  the  error function  ${\cal E}$ of the exact evolution operator $U_{num,2q}(T)$, computed numerically from $H_{2q}(t)$,  compared with the target gate $U_{tg}=U_{ent}$, as a function of the parameters $\omega$  and $A$.  In the  plot  of   Fig.\ref{fig:fidel2q}(a) 
we find a  point $A_{bS},\omega_{bS}$ with  a minimum  ${\cal E}$, near the numerical accuracy, which shows  that it is possible to implement the $\sqrt{\rm{bSWAP}}$  gate with this protocol.
To illustrate the  dynamical process that leads to  the $\sqrt{\rm{bSWAP}}$ gate, we show in Fig.\ref{fig:fidel2q}(b) the  time evolution of the population transfers during a driving period at $\omega_{bS},A_{bS}$.  
Furthermore, in Fig.\ref{fig:fidel2q}(c) we see that  in a non ideal case, when there is a small difference in the gaps of the qubits,  $\Delta_2=1.05\Delta_1$, the $\sqrt{\rm{bSWAP}}$ gate can be reproduced at a slightly shifted operational point and with error ${\cal E}\approx 10^{-6}$. 

We can proceed as in the previous section and provide an analytical estimate of the parameters for the $\sqrt{\rm{bSWAP}}$.
To use the CHRW approximation of Sec.\ref{s2} it is convenient to transform  $H_{2q}(t)$ to $\widetilde{H_{2q}}=SH_{2q}S^\dagger$ with 
$$S=\left(\begin{array}{cccc}
	1 & 0 & 0 & 0\\
	0 &0 & 0 & 1\\
	0 & 0 & 1 & 0\\
	0 & 1 & 0 & 0
\end{array}\right),$$

obtaining
\begin{equation}
\widetilde{H_{2q}}(t)=-\frac{1}{2}\left(\begin{array}{cccc}
	\Delta_{1}+\Delta_{2} & \epsilon(t) & 0 & 0\\
	\epsilon(t) & -\Delta_{1}-\Delta_{2} & 0 & 0\\
	0 & 0 & \Delta_{2}-\Delta_{1} & \epsilon(t)\\
	0 & 0 & \epsilon(t) & \Delta_{1}-\Delta_{2}
\end{array}\right),\nonumber
\end{equation}
which separates in two independent blocks,

\begin{eqnarray}
\widetilde{H_{2q}}(t)&=&\left[-\frac{\Delta_{+}}{2}\hat{\sigma}_{z}-\frac{\epsilon(t)}{2}\hat{\sigma}_{x}\right]\otimes\left(\begin{array}{cc}
	1 & 0\\
	0 & 0
\end{array}\right)\label{eq:h2qp}\\
&+&\left[-\frac{\Delta_{-}}{2}\hat{\sigma}_{z}-\frac{\epsilon(t)}{2}\hat{\sigma}_{x}\right]\otimes\left(\begin{array}{cc}
	0 & 0\\
	0 & 1
\end{array}\right),\nonumber
\end{eqnarray}
with $\Delta_{+}= \Delta_{1}+\Delta_{2}$
and $\Delta_{-}=\Delta_{2}-\Delta_{1}$.
Thus, in this basis we can express the evolution operator $\widetilde{U_{2q}}$ in terms of single qubit operators as,
\begin{equation}
    \widetilde{U_{2q}}(t)=U_+(t)\otimes\left(\begin{array}{cc}
	1 & 0\\
	0 & 0
\end{array}\right)
+U_-(t)\otimes\left(\begin{array}{cc}
	0 & 0\\
	0 & 1
\end{array}\right), 
\label{eq:U2qeff}
\end{equation}
where $U_+$  ($U_-$) is equal  to the single qubit evolution operator after replacing $\Delta$ by $\Delta_+$ ($\Delta_-$).
It is now straightforward to obtain  the evolution operator in the CHRW approximation, following the same steps as in the single qubit case for $U_+$ and $U_-$.
After one period $T$, and transforming back to the original basis, we obtain
\begin{equation}
\label{eq:u2q}
U_{2q}(T)=-\left(\begin{array}{cccc}
	a_+ & 0 & 0 & b_+\\
	0 & a_-^* & b_- & 0\\
	0 & -b_- & a_- & 0\\
    -b_+ & 0 & 0 & a_+^*
\end{array}\right),
\end{equation}
with
\begin{eqnarray}
a_+&=&  \cos\alpha_{+}+i\sin\alpha_{+}\cos{\widetilde\theta}_{+} \nonumber\\
b_+&=&\sin\alpha_{+}\sin{\widetilde\theta}_{+}\nonumber\\
a_-&=&\cos\alpha_{-}+i\sin\alpha_{-}\cos{\widetilde\theta}_{-} \nonumber\\
b_-&=&\sin\alpha_{-}\sin{\widetilde\theta}_{-}\nonumber,
\end{eqnarray}
where $\alpha_{+(-)}, {\widetilde\theta}_{+(-)}$ are obtained  as in Sec.\ref{s2}  after replacing $\Delta \rightarrow \Delta_{+} (\Delta_{-})$ in the generalized Rabi frequency $\Omega_R \rightarrow \Omega_{+R}(\Omega_{-R})$ and $(..)^*$ denotes the complex conjugate operation.

When $\Delta_{\text{1}}$=$\Delta_{2}=\Delta$, since $\Delta_{-}=0$ , we have
\begin{equation}
U_{2q}(T)=\left(\begin{array}{cccc}
	-a_+ & 0 & 0 & -b_+\\
	0 & 1 & 0 & 0\\
	0 & 0 & 1 & 0\\
	b_+ & 0 & 0 & -a_+^*
\end{array}\right).
\end{equation}



Therefore,  to obtain the  $\sqrt{\rm{bSWAP}}$ gate, a  comparison with Eq.(\ref{eq:uent}) gives   $b_+=-a_+=\frac{\sqrt{2}}{2}$ with $\Delta_+=2\Delta$ and the main the conditions are:

\begin{eqnarray}
\omega&=&\frac{4}{3}\Omega_{+R},
\\
\theta_+-\xi_+\frac{A}{\omega}&=&\frac{\pi}{2}
\nonumber
\end{eqnarray}

Numerical solution of these equations give  $\omega_{bS}=4.14\Delta$, $A_{bS}=5.74\Delta$, which coincide with the optimal point of Fig.\ref{fig:fidel2q}(a). (Note that they are the same as for the 
$Y_{\frac{\pi}{2}}$ gate after the substitution $\Delta \rightarrow 2\Delta$, an so similarly the general conditions  can be obtained.)

As in the single qubit case, for the  $\sqrt{\rm{bSWAP}}$ gate  we can extend the set of parameters to those  $A,\omega$ which satisfy $P_{00\rightarrow11}=1/2$. It is straightforward to show that this set is obtained from the solution of:
\begin{eqnarray}
P_{00\rightarrow11}&=&|\langle 11| U_{2q}(T)|00\rangle|{{}^2}\nonumber\\
&=&\sin{{}^2}{\tilde\theta_+}\sin{{}^2}\alpha_+=\frac{1}{2}
\label{eqP0011}.\\
\end{eqnarray}

To implement the $\sqrt{\rm{bSWAP}}$ gate for the parameters satisfying the above equation, one has to
add an idle time $t_{i}$ before the sinusoidal
drive and second idle time $t_{f}$ afterwards. Calling $\tau_{+}=(t_{1}+t_{2})\Delta$
, $\tau_{-}=(t_{2}-t_{1})\Delta$,
we have

\begin{equation}
U(t_{i}+T+t_{f})=e^{i\frac{\tau_{2}}{2}(\hat{\sigma}_{z}\otimes I+I\otimes\hat{\sigma}_{z})}U(T)e^{i\frac{\tau_{1}}{2}(\hat{\sigma}_{z}\otimes I+I\otimes\hat{\sigma}_{z})}
\end{equation}

\begin{equation}
=\left(\begin{array}{cccc}
	-e^{i\tau_{+}}a_+ & 0 & 0 & -e^{i\tau_{-}}b_+\\
	0 & 1 & 0 & 0\\
	0 & 0 & 1 & 0\\
	e^{-i\tau_{-}}b_+ & 0 & 0 & -e^{-i\tau_{+}}a_+^*
\end{array}\right).
\end{equation}

Defining $\rho e^{i\nu}=\cos\alpha_++i\sin\alpha_+\cos{\tilde\theta_+}$,
where $\tan\nu=\tan\alpha_+\cos{\tilde\theta_+}$, we
can write 

\begin{equation}
\left(\begin{array}{cccc}
	-\rho e^{i(\tau_{+}+\nu)} & 0 & 0 & \mp e^{i\tau_{-}}\sqrt{1-\rho^{2}}\\
	0 & 1 & 0 & 0\\
	0 & 0 & 1 & 0\\
	\pm e^{-i\tau_{-}}\sqrt{1-\rho^{2}} & 0 & 0 & -\rho e^{-i(\tau_{+}+\nu)}
\end{array}\right).
\end{equation}

For the case $P_{00\rightarrow11}=1-\rho^{2}=1/2$ this corresponds
to

\begin{equation}
\left(\begin{array}{cccc}
	\frac{\sqrt{2}}{2}e^{i(\tau_{+}+\nu)} & 0 & 0 & \mp\frac{\sqrt{2}}{2}e^{i\tau_{-}}\\
	0 & 1 & 0 & 0\\
	0 & 0 & 1 & 0\\
	\pm\frac{\sqrt{2}}{2}e^{-i\tau_{-}} & 0 & 0 & \frac{\sqrt{2}}{2}e^{-i(\tau_{+}+\nu)}
\end{array}\right).
\end{equation}

Then the $\sqrt{\rm{bSWAP}}$ gate can be obtained for $\tau_{-}=2k\pi$ and
$\tau_{+}+\nu=(2n+1)\pi$.

This gate is robust against a small difference in the parameters of the two qubits. For  $\Delta_2-\Delta_1=\epsilon\Delta$,
the invariance  under gate operation of the subspace spanned by $\{|01\rangle,|10\rangle\}$ can not be attained exactly.
Then, the error in the gate operation can be estimated from evaluating the probability $P_{|10\rangle\rightarrow|01\rangle}$, which should be zero for a perfect gate.
For $\epsilon\ll 1$ we estimate the error from Eq.(\ref{eq:u2q}) as ${\cal E}_{\epsilon} \propto  P_{|10\rangle\rightarrow|01\rangle}= b_-^2\approx\epsilon^2\pi^2(\frac{\Delta}{\omega})^2 J_0^2(\frac{A}{\omega})\sin^2(\frac{A}{\omega})$. For the case $(A_{bS},\omega_{bS}) $ and $\epsilon=0.01$ the error is ${\cal E}_{0.01}  \approx 2\times 10^{-5}$ and it decreases as $\sim \omega^{-2}$ for increasing $\omega$.


\section{Relaxation and decoherence under strong drive}\label{sec:dr}

In the previous sections we have found more than one  choice for the operational parameters $\omega,A$ to implement single qubit and two qubit gates with a LZSM protocol. 
In this section we analyze the effects of the environment on the gate dynamics since it is known that for strong driving the transition rates can depend on the driving parameters \cite{kohler_1998, hausinger_2010, Ferron_2012, Ferron_2016,yan_2013,yoshihara_2014}. Therefore  the dependence of  relaxation and decoherence rates  on  $\omega,A$ has to be considered to fine tune the implementation of qubit gates under these protocols. We will discuss here  the single qubit case,  but the analysis can be extended straightforwardly for the case  of  two qubits  considering that the dynamics of the Hamiltonian of Eq.(\ref{eq:h2q}), can be transformed to the dynamics of two independent qubits  as shown in Eqs.(\ref{eq:h2qp}) and (\ref{eq:U2qeff}).

The effect of the environment can be described by the global Hamiltonian ${\cal H}(t)={H}_{s}(t) + {H}_{b} + {H}_{sb}$,
where ${H}_{s}(t)={H}_{s}(t+T)$ is the Hamiltonian of the driven  qubits  with time period $T=2\pi/\omega$.
The Hamiltonian ${H}_{b}$ corresponds to a  thermal bath and 
${H}_{sb}=\hat{O}\otimes\hat{B}$ is the system-bath coupling term, with  $\hat{B}$ representing the quantum noise due to the bath and $\hat{O}$ is the observable of the system coupled to the noise.  

\begin{figure}[ht]
	\centering
	\includegraphics[width=\linewidth]{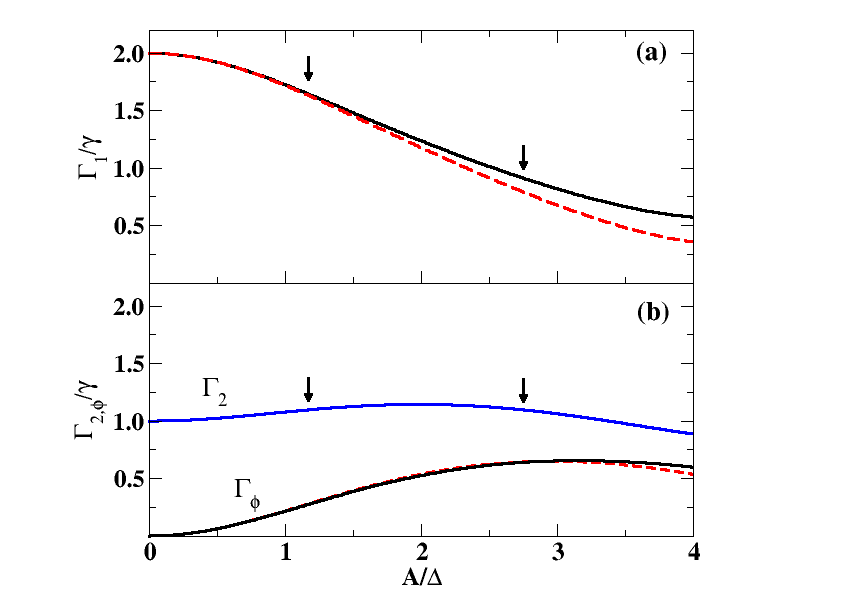}
	\caption{ (a) Plot of the relaxation rate $\Gamma_1$ as a function of $A/\Delta$  for $\omega=1.92\Delta$.
The black continuous line corresponds to the numerically exact value and the dashed line the CHRW approximation,  Eq.(\ref{eq:gamma1a}). (b)
Plot of the dephasing rate $\Gamma_\phi$ as a function of $A/\Delta$  for $\omega=1.92\Delta$.
The black continuous line corresponds to the numerically exact value and the dashed line the CHRW approximation. 
Also the decoherence rate $\Gamma_2$ is plotted (blue continuous line, numerically exact values only).
Both in (a) and (b): the arrows indicate the values of $A$ where  the $X_{\frac{\pi}{2}}$, $Y_{\frac{\pi}{2}}$ gates can be implemented, as obtained from Fig.\ref{fig:figsin}(a); the rates $\Gamma_1,\Gamma_\phi,\Gamma_2$ are normalized by the noise strength parameter $\gamma$ and correspond to a thermal bath at temperature $T_b=0.1\Delta$.}
 \label{fig:rates}
\end{figure}

The  natural basis to compute relaxation and decoherence rates in the case
of strong time periodic drives is the Floquet basis \cite{Shirley_1965}, since in this basis the density matrix in the steady state becomes diagonal \cite{grifoni_1998,grifoni_2010,kohler_1997,kohler_1998,breuer_2000,hone_2009,Ferron_2016,gasparinetti_2013,gasparinetti_2014}. In the case of the two level system, like the Hamiltonian of Eq.(\ref{eq:hamil}), the  wave functions have the time dependence $|\Psi(t)\rangle=c_ae^{-i\epsilon_a t}|a(t)\rangle + c_b e^{-i\epsilon_b t}|b(t)\rangle$, where the  Floquet states $|a(t)\rangle$, $|b(t)\rangle$ are periodic  with time period $T$, and $\epsilon_a$, $\epsilon_b$  are the associated quasienergies \cite{ Shirley_1965, grifoni_1998,grifoni_2010,ferron_2010}. In the CHRW approximation, $|a(t)\rangle$ and $|b(t)\rangle$ can be obtained from the eigenstates of the static Hamiltonian, Eq.(\ref{eqnhdtilde}), after performing on them a time dependent transformation back to the representation of the original Hamiltonian Eq.(\ref{eq:hamil}).
In the  limit $A\rightarrow0$ the Floquet states tend to the  eigenstates of the undriven Hamiltonian: $|a(t)\rangle\rightarrow|0\rangle$, $|b(t)\rangle\rightarrow|1\rangle$, and similarly the Floquet gap
$\Delta_F= |\epsilon_b-\epsilon_a |$ tends to the undriven gap, $\Delta_F\rightarrow \Delta=E_1-E_0$.

From the Floquet-Markov quantum master equation (see Appendix A) the relaxation rate can be obtained  as,
\begin{eqnarray}
   \Gamma_{1}&=&\sum_{q}S(\epsilon_{b}-\epsilon_a+q\omega)
\left|\frac{1}{T}\int_0^T\langle a(t)|\hat{O}|b(t)\rangle e^{iq\omega t}dt\right|^{2}\nonumber\\
&\approx& S(\Delta_F)\left|\frac{1}{T}\int_0^T\langle a(t)|\hat{O}|b(t)\rangle dt\right|^{2},
\label{eq:gamma1}
\end{eqnarray}
where $S(\Omega)$ is the noise power spectrum. The second line of Eq.(\ref{eq:gamma1}) approximates $\Gamma_1$  with the $q=0$ term, which is the dominant contribution in the sum of the first line. It is also direct to show that in the undriven limit, $A\rightarrow0$, we can recover the standard result $\Gamma_{1}=S(\Delta)|\langle 0|\hat{O}|1\rangle|^{2}$.

 We consider here that the main source of quantum noise is through the same channel as the driving, and thus we take for the noise coupling operator $\hat{O}=\hat{\sigma}_x$.
An approximate expression  of $\Gamma_1$ can be obtained 
calculating the matrix elements $\langle a(t)|\hat{\sigma}_x|b(t)\rangle $ in  the CHRW approximation (see Appendix A),  
\begin{equation}
    \Gamma_1\approx S(\Delta_F)\cos^4\frac{\theta}{2}.
    \label{eq:gamma1a}
\end{equation}
where $\Delta_F=|\Omega_R-\omega|$ in this case.
We have calculated the dependence of $\Gamma_1$ with the frequency $\omega$ and the amplitude $A$ considering 
 noise with power spectrum $S(\Omega)=2\gamma\Omega\coth({\Omega}/{2T_b})$.
 We obtain the values of $\Gamma_1$ normalized by the noise strength parameter $\gamma$. In order to see more clearly the dependence with the driving parameters, we show the case for a low
temperature $T_b=0.1\Delta$.
We find that the general behavior for $\omega>\Delta$ is that the relaxation rate decreases with increasing $A$, as seen  in  Fig.\ref{fig:rates}(a) for frequency $\omega=1.96\Delta$ in the range of interest \cite{comment}.  Higher temperatures ($T_b\gtrsim\Delta$) give a similar behavior but with a milder dependence with $A$.
The arrows in Fig.\ref{fig:rates}(a) indicate the values of $A$ for which the $X_{\frac{\pi}{2}}$, $Y_{\frac{\pi}{2}}$ gates could be implemented at this driving frequency, as obtained from Fig.\ref{fig:figsin}(a). Considering that $\Gamma_1$ is smaller for larger $A$, the value indicated by the second arrow in the plot should be the preferred  choice for a reduced effect of the environment in the qubit dynamics.
 We also plot in  Fig.\ref{fig:rates}(a)  the CHRW approximation of Eq.(\ref{eq:gamma1a}) and the numerically exact evaluation of Eq.(\ref{eq:gamma1}) 
(after calculating  the Floquet states and quasienergies and summing terms in $q$ up $\pm 32$), showing that they
are  in good agreement.

To complete the analysis of the effect of the environment, we have to calculate the decoherence rate $\Gamma_{2}={\Gamma_{1}}/{2}+\Gamma_\phi$.  The dephasing rate $\Gamma_\phi$ can be obtained from the Floquet-Markov quantum master equation as
\begin{equation}
\Gamma_\phi= \sum_{q\ge0}2S(q\omega) \left|\frac{1}{T}\int_0^Tdt\langle a(t)|\hat{O}|a(t)\rangle e^{iq\omega t}\right|^{2}.
\end{equation}
In the case under consideration, with noise coupling operator $\hat{O}=\hat{\sigma}_x$, the $q=0$ term is exactly zero.
Moreover, in the undriven limit $A\rightarrow0$ the dephasing  rate completely vanishes, $\Gamma_\phi^{A\rightarrow0}=0$, corresponding to the fact that the qubit of Eq.(\ref{eq:hamil}) is in a "sweet spot" \cite{campbell_2020}.
But, for finite driving, the $q\not=0$ terms start to contribute to dephasing with the dominant term being the $q=1$ term, leading to the expression
\begin{equation}
\Gamma_\phi\approx 2S(\omega) \left|\frac{1}{T}\int_0^Tdt\langle a(t)|\hat{\sigma}_x|a(t)\rangle e^{i\omega t}\right|^{2}.
\end{equation}
In this case, the CHRW approximation gives
\begin{equation}
    \Gamma_\phi\approx 2S(\omega)[\sin\theta(J_{0}+J_{2}) +\sin^2\frac{\theta}{2}(J_{1}+J_{3})]^2,
\end{equation}
where we have denoted $J_l\equiv J_l(\xi A/\omega)$.
We plot in Fig.\ref{fig:rates}(b) the dephasing rate $\Gamma_\phi$ as a function of $A$ for $\omega=1.96\Delta$. As stated, we find that $\Gamma_\phi=0$ for $A=0$,  and then  that $\Gamma_\phi$ increases for increasing $A$. Therefore dephasing is increased by the driving, which is in the opposite direction as the effect of driving on the relaxation rate, analyzed in the previous paragraph. However, to determine the optimal parameters for the gate, one has to analyze the decoherence rate $\Gamma_2$, that combines dephasing and relaxation. As can be seen in  Fig.\ref{fig:rates}(b),  the  decoherence rate changes mildly as a function of the driving strength $A$, being nearly the same for the two cases indicated by the arrows. Therefore, considering the previously discussed driving effect on relaxation, the larger $A$ is still the better choice for the implementation of the gates.
 We also compare in  Fig.\ref{fig:rates}(b)  the approximated and   the numerically exact $\Gamma_\phi$, which are in good agreement. 

 \begin{figure}[ht]
	\centering
	\includegraphics[width=\linewidth]{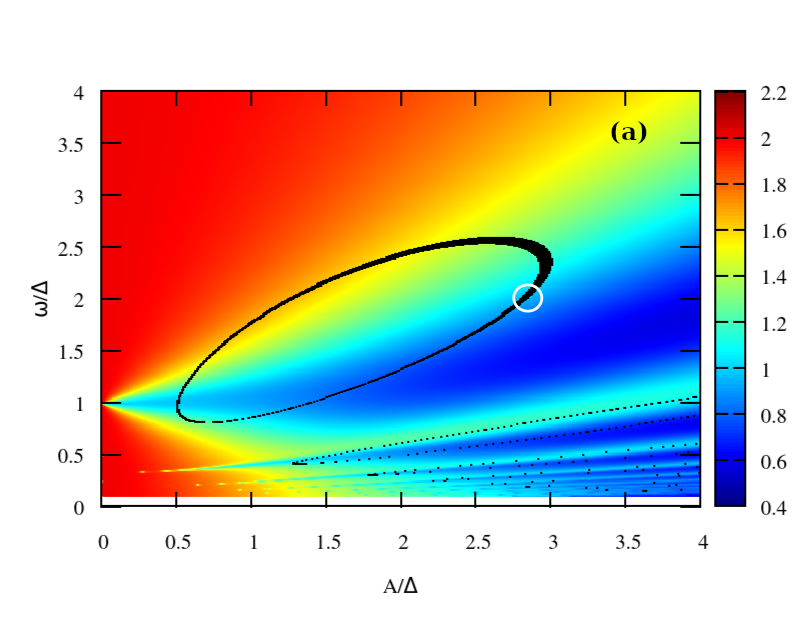}
	\includegraphics[width=\linewidth]{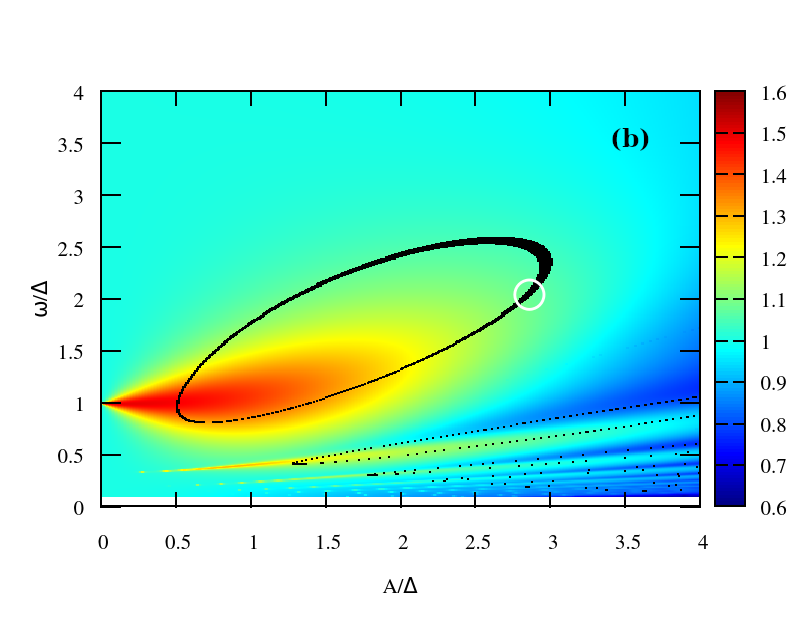}
	\caption{ Intensity plot of (a) the relaxation rate $\Gamma_1$  and (b) the   decoherence rate $\Gamma_2$, as a function of $A/\Delta$ and $\omega/\Delta$.  The black dots  indicate the parameter sets  $\omega$  and $A$ for the implementation of $X_{\frac{\pi}{2}}$ and $Y_{\frac{\pi}{2}}$ gates, as in Fig.\ref{fig:figsin}(a). The white circle shows the optimal parameter region to minimize environmental effects.
 The rates $\Gamma_1,\Gamma_2$ are normalized by the noise strength parameter $\gamma$ and correspond to a thermal bath at temperature $T_b=0.1\Delta$. }
 \label{fig:rates2}
\end{figure}

When considering lower  $\omega$ the behavior of $\Gamma_1$ and $\Gamma_2$ is more complex. In Fig.\ref{fig:rates2}(a) and (b) we show intensity plots of $\Gamma_1$ and $\Gamma_2$, respectively, as a function of $A$ and $\omega$, plotting the numerically exact values in the full range of $\omega$ (the approximated values discussed above are accurate only for $\omega>\Delta$). We also plot in Fig.\ref{fig:rates2} (with black dots) the values of $A,\omega$ corresponding to the conditions for the implementation of the $X_{\frac{\pi}{2}}$, $Y_{\frac{\pi}{2}}$ gates. The relaxation rate $\Gamma_1$ decreases for increasing $A$ for any of the frequencies in the range of interest ($\omega\gtrsim\Delta$), thus large $A$ would be always more convenient for gate implementations in order to minimize relaxation. In the plot of  Fig.\ref{fig:rates2}(a) this corresponds to the gate parameters that fall within the blue colored region (which indicates lowest values of $\Gamma_1$ in the color scale of the plot). 

On the other hand, the decoherence rate $\Gamma_2$ is large in the regions near  resonance $\omega\sim\Delta$ and for $A\lesssim2\Delta$, within the red colored region in Fig.\ref{fig:rates2}(b).
This behavior is almost independent of temperature, {\it i.e.},  higher temperatures ($T_b\gtrsim\Delta$) give similar plots except for a larger overall value of $\Gamma_2$, because the relevant dependence is on the matrix element $\left|\frac{1}{T}\int_0^Tdt\langle a(t)|\hat{\sigma}_x|a(t)\rangle e^{i\omega t}\right|^{2}$.
Therefore, off-resonant large frequency  driving is always more convenient for the gates analyzed here, since for $\omega\gg\Delta$ the decoherence rate is low and nearly insensitive to variations in the driving amplitude $A$.

The  combined analysis of the competing conditions for minimal relaxation and   minimal decoherence lead to the conclusion that the best parameters for the implementation of the $X_{\frac{\pi}{2}}$, $Y_{\frac{\pi}{2}}$ gates are within the  region in ${A,\omega}$ highlighted with a circle in Fig.\ref{fig:rates2}.

\section{Summary and Conclusions}\label{sec:concl}

We have analyzed LZSM strong driving protocols for the implementation of quantum gates which are well suited for small gap qubits. We provide accurate analytical equations to obtain the driving parameters (amplitude, frequency, initial and final idling times) for single qubit gates and for the $\sqrt{\rm{bSWAP}}$ two qubits gate. Our approach avoids the need to  scan experimentally a wide range of parameters and instead it allows to focus in fine-tuning the device near the analytically predicted parameters.

We have found that the $X_{\frac{\pi}{2}}$ and $Y_{\frac{\pi}{2}}$ gates can be efficiently implemented with a single strong one-period sinusoidal drive, with parameters in the range $\Delta \lesssim \omega \lesssim 3\Delta$ and  $\Delta \lesssim A \lesssim 3\Delta$. We note that the $X_{\frac{\pi}{2}}$ and $Y_{\frac{\pi}{2}}$ gates could also be implemented using a   half-period sinusoidal drive, which would allow  for operation at even larger amplitudes and frequencies (it is easy  in the CHRW calculation to obtain the $U(T/2)$ operator and the corresponding conditions for the gates). However, a one-period sinusoidal drive is preferred since it has zero time integral and thus the  dc components associated with pulse transients  cancel out \cite{campbell_2020}.

The high amplitude and high frequency of the sinusoidal pulses make necessary 
to take into account the dependence of relaxation and decoherence with the driving parameters.
We have shown that relaxation and decoherence decrease with increasing amplitude. Therefore large $A$ should be preferred.
However, leakage to higher energy levels could induce gate errors for large drives. This effect depends on the specific multilevel
structure of the quantum device. A rule of thumb argument is that the   amplitude $A$ should be smaller than $E_2-E_1$, with $E_2$ the energy of the
third  level (and $E_0,E_1$ the qubit two-level energies), to avoid leakage effects. In the optimal region signaled in Fig.\ref{fig:rates2} we find $A\sim 3\Delta$ for minimal relaxation and decoherence, then requiring  $E_2-E_1 \gg 3(E_1-E_0)$.  Most superconducting qubit devices fulfill this condition.  Decoherence is much smaller in the off-resonant case, for frequencies $\omega>\Delta$.  After one driving period $T$ the error due to relaxation is proportional to $\Gamma_1T=2\pi\Gamma_1/\omega$ and similarly the error due to decoherence is proportional to $2\pi\Gamma_2/\omega$. Therefore, high frequencies,  which imply faster gates, are always preferred to reduce the detrimental effects of the environment. 

Here we propose the $\sqrt{\rm{bSWAP}}$ gate as the simplest two-qubit gate that can be implemented with  a strong driving LZSM protocol. Previous implementations of the $\sqrt{\rm{bSWAP}}$ gate have been with protocols based on two-photon transitions \cite{poletto_2012,roth2017,nesterov2021}.
The protocol based on LZSM transitions proposed here  only requires a single one period sinusoidal pulse, and thus it can be easier to realize, and possibly faster, than  the ``two-photon" protocols. Therefore, we consider to be worthwhile to implement in the future this  two-qubit gate   in  small gap superconducting qubits.

\section*{Acknowlegments}
We acknowledge support from CNEA, CONICET , ANPCyT ( PICT2019-0654) and  UNCuyo (06/C591).

\appendix

\section{Floquet states and quantum master equation}
\subsection{Floquet states}
Consider the two level Hamiltonian:

\begin{equation}
	H(t)=-\frac{\Delta}{2}\hat{\sigma}_{z}-\frac{\epsilon(t)}{2}\hat{\sigma}_{x}
 \label{eqhamilA}
\end{equation}
with $\epsilon(t)=A\sin(\omega t)$.

According to Floquet theorem for time-periodic Hamiltonians, the solutions of the  Schr\"odinger equation are of the form $|\Psi_\alpha(t)\rangle=e^{i\epsilon_\alpha t/\hbar}|u_\alpha(t)\rangle$, where the  Floquet states $|u_\alpha(t)\rangle$ satisfy $|u_\alpha(t)\rangle$=$|u_\alpha(t+ T)\rangle$
and are eigenstates of 
$[{H}_s (t)- i \hbar \partial/\partial t ] |u_\alpha(t)\rangle= \epsilon_\alpha |u_\alpha(t)\rangle$, with $\epsilon_\alpha$ the associated quasienergy \cite{Shirley_1965,grifoni_1998,grifoni_2010}. 
The evolution operator can be written in matrix form as

\[
U(t)=P(t)e^{-iH_{F}t}P^{\dagger}(0)
\]
where $P(t+T)=P(t)$ is the matrix that contains the components of the Floquet states (in a given basis). 
In the CHRW approximation we obtain for the matrix of Floquet states
\[
P(t)=U_{x}^{\dagger}(t)R^{\dagger}(t)W^{\dagger}e^{i\frac{\omega t}{2}\hat{\sigma}_{z}}
\]
where  we have taken for $H_F$ the form:
\[
H_{F}=-\frac{\Omega_{R}-\omega}{2}\hat{\sigma}_{z}
\]
which gives  the correct $\omega\rightarrow\infty$ limit. 

From the columns of the $P(t)$ matrix we obtain the Floquet states

\[
\left|u_\alpha(t)\right\rangle =\left\{ \begin{array}{c}
\left|a\right\rangle =\left(\begin{array}{c}
\cos\frac{\theta}{2}\cos\frac{\phi}{2}-e^{-i\omega t}\sin\frac{\theta}{2}\sin\frac{\phi}{2}\\
i\cos\frac{\theta}{2}\sin\frac{\phi}{2}+ie^{i\omega t}\sin\frac{\theta}{2}\cos\frac{\phi}{2}
\end{array}\right)\\
\\
\left|b\right\rangle =\left(\begin{array}{c}
i\cos\frac{\theta}{2}\sin\frac{\phi}{2}+ie^{-i\omega t}\sin\frac{\theta}{2}\cos\frac{\phi}{2}\\
\cos\frac{\theta}{2}\cos\frac{\phi}{2}-e^{i\omega t}\sin\frac{\theta}{2}\sin\frac{\phi}{2}
\end{array}\right)
\end{array}\right.,
\]
where $\theta=-\arctan{\widetilde{A}/\widetilde{\delta}}$ and 
 $\phi(t)=-\xi\frac{A}{\omega}\cos(\omega t)$ were already defined in Sec.\ref{s2}.
The corresponding quasienergies are

\[
\epsilon_{a/b}=\mp\frac{\Omega_{R}-\omega}{2},
\]
and the so called Floquet gap is $\Delta_F=|\epsilon_{b}-\epsilon_{a}|=|\Omega_{R}-\omega|$.
%






\subsection{Floquet-Markov master equation and transition rates}

 The open system dynamics can be described by the global Hamiltonian ${\cal H}(t)={H}_{s}(t) + {H}_{b} + {H}_{sb}$,
where ${H}_{s}(t)={H}_{s}(t+T)$ is the Hamiltonian of the qubits driven by periodic external fields with time period $T=2\pi/\omega$.
The Hamiltonian ${H}_{b}$ corresponds to a bosonic thermal bath at temperature $T_{b}$ and spectral density $J(\Omega)$. The bath  h is linearly coupled to  the qubit system in the form 
${H}_{sb}=\hat{O}\otimes\hat{B}$, with  $\hat{B}$ an observable of the bath and $\hat{O}$ an observable of the system.  
After performing the Born and Markov approximations, a quantum master equation can be obtained \cite{grifoni_1998,grifoni_2010,kohler_1997,kohler_1998,breuer_2000,hone_2009,Ferron_2016,gasparinetti_2013,gasparinetti_2014}. 
 In most situations (away from resonances) an additional secular approximation can be realized \cite{grifoni_1998,grifoni_2010,kohler_1997,kohler_1998,breuer_2000,hone_2009,Ferron_2016,gasparinetti_2013,gasparinetti_2014,gramajo_2018}, leading to the   quantum master equation:
\begin{equation}
	\begin{aligned}
	\dot{\rho}&= -i [ H_{s}(t),\rho] + \sum_{\alpha\beta}  \Gamma_{\alpha\beta} \left( L_{\alpha\beta}\rho L^{\dag}_{\alpha\beta} - \frac{1}{2} \{ L^{\dag}_{\alpha\beta}L_{\alpha\beta},\rho \}\right),
	\label{eq:Linblad}
	\end{aligned}
\end{equation} where $L_{\alpha\beta} = |u_{\alpha}(t)\rangle \langle u_{\beta}(t)|$ are the corresponding jump operators, and the transition rates $\Gamma_{\alpha\beta}$ can be written as 
\begin{equation}	
	\Gamma_{\alpha\beta} = \sum_{q}  g(\epsilon_{\alpha\beta,q})|O_{\alpha\beta}(q)|^{2},
\end{equation} 
where  the $q$-Fourier components of the transition matrix elements are
$$O_{\alpha\beta}(q)=\frac{1}{T}\int_0^T dt \langle u_\alpha(t)|\hat{O}|u_\beta(t)\rangle  e^{iq\omega t},$$
and $g(x)$ is  the spectral bath correlation function, $g(x)=J(x)n_{\rm th}(x)$ with
with $J(x)$ the spectral density, $n_{\rm th}(x)=(\exp{(x/k_BT)}-1)^{-1}$, and
 $\varepsilon_{\alpha\beta,q}=\varepsilon_\alpha-\varepsilon_{\beta}+q\hbar \omega$.


In the case of a two-level system like the Hamiltonian of Eq.(\ref{eqhamilA}) the relaxation rate can be obtained from the Eq.(\ref{eq:Linblad}) as

$$
\Gamma_{1}=\sum_{q}g(\epsilon_{ab,q})|O_{ab}(q)|^{2}+g(\epsilon_{ba,q})|O_{ba}(q)|^{2}$$
Using that $O_{ab}(q)=\left[O_{ba}(-q)\right]^{*}$  we can write
$$
\Gamma_{1}=\sum_{q}S(\epsilon_{ab,q})|O_{ab}(q)|^{2}$$
where $S(\Omega)$ is the noise power spectrum, $S(\Omega)=g(\Omega)+g(-\Omega)$.
The decoherence rate is $\Gamma_{2}=\frac{\Gamma_{1}}{2}+\Gamma_\phi$ with the dephasing rate,
$$\Gamma_\phi= \sum_{q}g(q\omega) |O_{aa}(q)-O_{bb}(q)|^{2}$$ 
Without loss of generality we can choose ${\rm Tr}({\hat O})=0$, and then $O_{aa}(q)=-O_{bb}(q)$, giving
$$\Gamma_\phi= \sum_{q\ge0}2S(q\omega) |O_{aa}(q)|^{2}$$ 


 For $\hat{O}=\hat{\sigma}_{x}$ the matrix elements are $\langle u_\alpha(t)|\hat{O}|u_\beta(t)\rangle=\left\langle u_\alpha(t)|\hat{\sigma}_{x}|\beta(t)\right\rangle\equiv X_{\alpha\beta}(t)$.
In the CHRW approximation we obtain the expressions
 \begin{eqnarray}
    X_{ab}(t) &=&\cos{{}^2}\frac{\theta}{2}+\sin{{}^2}\frac{\theta}{2}\cos2\omega t\nonumber\\
    &&-i\left(\sin\omega t\sin\theta\sin\phi+\sin2\omega t\cos\phi\right)\nonumber
\end{eqnarray}
$$X_{aa}(t) =-\sin\omega t\sin\theta\cos\phi+\sin2\omega t\sin{{}^2}\frac{\theta}{2}\sin\phi$$

To evaluate the rates,  the $q$ Fourier components $X_{\alpha\beta}(q)$ have to be calculated. After using the expansions for $\sin\phi(t)$ and $\cos\phi(t)$ 
$$
\sin\phi(t)=\sum_{l}(-1)^{l+1}J_{2l+1}e^{i(2l+1)\omega t}
$$
$$
\cos\phi(t)=\sum_{l}(-1)^{l}J_{2l}e^{i2l\omega t}
$$
with $J_l\equiv J_l(\xi A/\omega)$, we have,
\begin{eqnarray}
    X_{ab}(2l+1)&=&0  \nonumber \\ 
    X_{ab}(2l)&=&\frac{(-1)^{l+1}}{2}\left[\sin\theta\left(J_{2l+1}+J_{2l-1}\right)\right.\nonumber\\
    & &\left. -J_{2l-2}+J_{2l+2}\right]\nonumber\\
    & &+\delta_{l,0}\cos{{}^2}\frac{\theta}{2}+\left(\frac{\delta_{l,1}+\delta_{l,-1}}{2}\right)\sin{{}^2}\frac{\theta}{2}
    \nonumber\\
    X_{aa}(2l+1)&=&\frac{i^{2l+1}}{2}\left[ \sin\theta\left(J_{2l}+J_{2l+2}\right)
    \right.\nonumber\\
    & & \left. -\sin{{}^2}\frac{\theta}{2}\left(J_{2l-1}-J_{2l+3}\right)\right]\nonumber\\
    X_{aa}(2l)&=&0\nonumber .
\end{eqnarray}
In the lowest approximation the relaxation rate is dominated by the $q=0$ term, giving
$$\Gamma_1\approx2S(\Delta_F)|X_{ab}(0)|^{2}=2S(\Delta_F)\cos^4\frac{\theta}{2},$$
where the noise spectrum is evaluated at the Floquet gap $\Delta_F=|\Omega_R-\omega|$.
On the other hand, for the dephasing rate, the $q=0$ term is zero, and we have to
take the next term as an approximation,
$$\Gamma_\phi\approx 2S(\omega) |X_{aa}(1)|^{2}=S(\omega)[(J_{0}+J_{2})\sin\theta +(J_{1}+J_{3}) \sin^2\frac{\theta}{2}]^2 $$

\section{Other approximation methods to the dynamics}

The dynamics of the driven two level system has been studied extensively along the last years. Different approximation methods have been attempted to solve  the dynamics of a strongly driven qubit, given by the Hamiltonian Eq.(\ref{eq:hamil})  for $\epsilon(t)= A\sin(\omega t)$.
Here we review some  and compare them with the CHRW approximation.

\subsection{Double rotating frame rotating wave approximation (DR)}

The dynamics can also be approximated following the approach of Refs.\cite{deng_2015,deng_2016} where  an improved  (``second order") rotating wave approximation is performed to calculate the Floquet states and quasienergies, after a basis transformation to a rotating frame with a time-dependent rotation frequency and a truncation of the transformed Floquet Hamiltonian to a 2 × 2 matrix. 

Here we obtain the same result following a different (but equivalent) procedure, where we perform   two rotation transformations of the Hamiltonian Eq.(\ref{eq:hamil}) and a RWA approximation at the end.
 We start with the $x$-rotation $|\psi'(t)\rangle=U_x|\psi(t)\rangle$, with $U_{x}=e^{-i\frac{\phi}{2}\hat{\sigma}_{x}}$, and  $\phi(t)=-\frac{A}{\omega}\cos(\omega t)$. After the rotation, the  transformed Hamiltonian is
 $H'=U_xHU_x^\dagger+i(\partial_t U_x)U_x^\dagger$ and thus, 
\begin{equation}
	H'=-\frac{\Delta}{2}(\cos\phi \hat{\sigma}_{z}-\sin\phi \hat{\sigma_{y}}).
 \label{eha1}
\end{equation}

Using the  expansion of $e^{i\phi(t)}$ in terms of Bessel functions 
we approximate in Eq.(\ref{eha1}) (neglecting the high frequency terms):
\begin{eqnarray}
\cos\phi&\approx& J_{0}(\frac{A}{\omega})\nonumber\\
\sin\phi&\approx&-2J_{1}(\frac{A}{\omega})\cos(\omega t),
\end{eqnarray}

and therefore
\begin{equation*}
H'=-\frac{\Delta}{2}\left[J_{0}(\frac{A}{\omega})\hat{\sigma}_{z}+2J_{1}(\frac{A}{\omega})\cos(\omega t)\hat{\sigma_{y}}\right].
\end{equation*}


The second rotation is a $z$-rotation with the unitary operator $U_z=e^{-i\frac{\omega t}{2}\hat{\sigma}_{z}}$,  for which we obtain:

\begin{equation}
H'' 
=-\frac{\widetilde{\delta}}{2}\hat{\sigma}_{z}+\frac{\widetilde{\nu}}{2}\left[(1+\cos2\omega t)\hat{\sigma_{y}}-\sin2\omega t \hat{\sigma_{x}}\right],
\end{equation}
being $\widetilde{\delta}=\Delta J_{0}(a)-\omega$, $\widetilde{\nu}=\Delta J_{1}(\frac{A}{\omega})$ and $a= A/ \omega$. Neglecting the fast oscillating terms with frequency $2\omega$ (RWA approximation),

\begin{eqnarray}
H''\approx -\frac{\widetilde{\delta}}{2}\hat{\sigma}_{z}+\frac{\widetilde{\nu}}{2}\hat{\sigma_{y}}.
\label{eqnhdtildea}
\end{eqnarray}
Equation (\ref{eqnhdtildea}) can be easily  diagonalized with the transformation $W=e^{-i\frac{\theta}{2}\hat{\sigma}_{x}}$, being
$\tan\theta=-\widetilde{\nu}/\widetilde{\delta}$, obtaining:

\begin{equation}
H_{d}=WH''W^{\dagger}=-\frac{\Omega_{R}}{2}\hat{\sigma}_{z},
\end{equation}

with 
$$\Omega_{R}=\sqrt{\widetilde{\delta}{{}^2}+\widetilde{\nu}{{}^2}}=\sqrt{[\Delta J_{0}(\frac{A}{\omega})-\omega]^{2}+\Delta^{2}J_{1}{{}^2}(\frac{A}{\omega})},$$

the generalized Rabi frequency. 

Taking into account the previous transformations,  the  evolution operator associated to Eq.(\ref{eq:hamil}) results :

\begin{equation}
U^{DR}(t)=U_{x}^{\dagger}(t)U_z^{\dagger}(t)W^{\dagger}e^{i\frac{\Omega_{R}t}{2}\hat{\sigma}_{z}}WU_z(0)U_{x}(0),
\end{equation}

which after one period of the driving, $T=2\pi/\omega$, is:

\begin{equation}
U^{DR}(T)=-e^{i\frac{\theta-a}{2}\hat{\sigma}_{x}}e^{i\frac{\pi\Omega_{R}}{\omega}\hat{\sigma}_{z}}e^{-i\frac{\theta-a}{2}\hat{\sigma}_{x}}
\end{equation}
\begin{equation*}
=-\left(\begin{array}{cc}
	\cos\alpha+i\sin\alpha\cos{\tilde\theta} & \sin\alpha\sin{\tilde\theta}\\
	-\sin\alpha\sin{\tilde\theta} & \cos\alpha-i\sin\alpha\cos{\tilde\theta}
\end{array}\right)\nonumber,
\end{equation*}

with $\alpha=\frac{\pi\Omega_{R}}{\omega}$ and  ${\tilde\theta}=\theta-\frac{A}{\omega}$.

We can now calculate in this approximation the transition probability between the qubit states, $\left|0\right\rangle \rightarrow\left|1\right\rangle $ after a time $t=T$ as,

\begin{equation}
    P_{01}^{DR}=|\langle 1| U(T)|0\rangle|{{}^2}=\sin{{}^2}{\tilde\theta}\sin{{}^2}\alpha
\label{eqP01ap}
\end{equation}
which is very similar in form to the obtained in the CHRW approximation. (Note that here the  frequency $\Omega_R$ and the angles $\alpha$,$\theta$, etc. have different expressions).





\subsection{Magnus expansion approximation (ME)}

In \cite{weiss_2022} the dynamics  is approximated with a Magnus expansion \cite{magnus_1954,blanes_2009}.
Considering a Hamiltonian $H(t)$, the Magnus expansion for the evolution operator $U(t_f,t_i)$ from time $t = t_i$ to time $t = t_f$, with  $\Delta t= t_f - t_i$, is

\begin{equation}
U\left(t_f, t_i\right)=\exp \{-\mathrm{i} \bar{H} \Delta t\},
\end{equation}
with
\begin{equation}
\bar{H}=\bar{H}^{(1)}+\bar{H}^{(2)}+\bar{H}^{(3)}+\bar{H}^{(4)}+\ldots\;,
\end{equation}
where the first terms $\bar{H}^{(n)}$ of the  expansion are:
\begin{equation}
\begin{aligned}
\bar{H}^{(1)}=&\frac{1}{\Delta t} \int_{t_i}^{t_f} \mathrm{~d} t H(t),\\
\bar{H}^{(2)}=& \frac{1}{2 \mathrm{i} \Delta t} \int_{t_i}^{t_f} \mathrm{~d} t \int_{t_i}^{t} \mathrm{~d} t^{\prime}\left[H(t), H\left(t^{\prime}\right)\right]. 
\end{aligned}
\label{magnus-expansion}
\end{equation}

Since for fast gates we are interested in the evolution after one period of the drive $T=2\pi/\omega$, the Magnus expansion can be used to estimate $U(T)$ for the  Hamiltonian  $H$ given in Eq.(\ref{eq:hamil}).
Following the approach of \cite{weiss_2022},  we start by applying  the transformations $|\psi'(t)\rangle=U_0|\psi(t)\rangle$,
 $H'=U_0HU_0^\dagger+i(\partial_t U_0)U_0^\dagger$, with $U_{0}=e^{-i\frac{\chi}{2}\hat{\sigma}_{x}}$,  and $\chi(t)=-\frac{A}{\omega}[\cos(\omega t)-1]$. This gives 
\begin{equation}
	H'=-\frac{\Delta}{2}(\cos\chi \hat{\sigma}_{z}-\sin\chi \hat{\sigma_{y}}).
\end{equation}
For the evolution after one period  $T=2\pi/\omega$ we consider the lowest order in the Magnus expansion:
$$U'(T) \approx e^{-i\frac{1}{T} \int_{0}^{T} \mathrm{~d} t H'(t)}.$$
The Magnus expansion converges for $\|H(t)\| \Delta t \ll 1$, which in this case corresponds to the high frequency limit $\Delta \ll \omega$. 
Thus one obtains \cite{weiss_2022}
$$U'(T) \approx 
-\left(\begin{array}{cc}
	\cos\alpha+i\sin\alpha\cos{\theta} & \sin\alpha\sin{\theta}\\
	-\sin\alpha\sin{\theta} & \cos\alpha-i\sin\alpha\cos{\theta}
\end{array}\right),
$$
with $\alpha=\pi\Omega_R^{(0)}/\omega$, $\theta=A/\omega$, and $\Omega_R^{(0)}=\Delta J_0(\frac{A}{\omega})$. Since $U_0(0)=U_0(T)=I$, the evolution operator is $U(T)=U_0(T)U´(T)U_0^\dagger(0)=U'(T)$. Therefore the transition probability $P_{01}$ in this first order Magnus expansion approximation is 
\begin{eqnarray}
P_{01}^{ME}&=&|\langle 1| U(T)|0\rangle|{{}^2}=\sin{{}^2}{\theta}\sin{{}^2}\alpha \label{eqP01MEA}\\
&=&\sin^2(\frac{A}{\omega})\sin{{}^2}\frac{\pi\Delta J_0(\frac{A}{\omega})}{\omega}.\nonumber
\end{eqnarray}

\begin{figure}[ht]
	\centering
	\includegraphics[width=\linewidth]{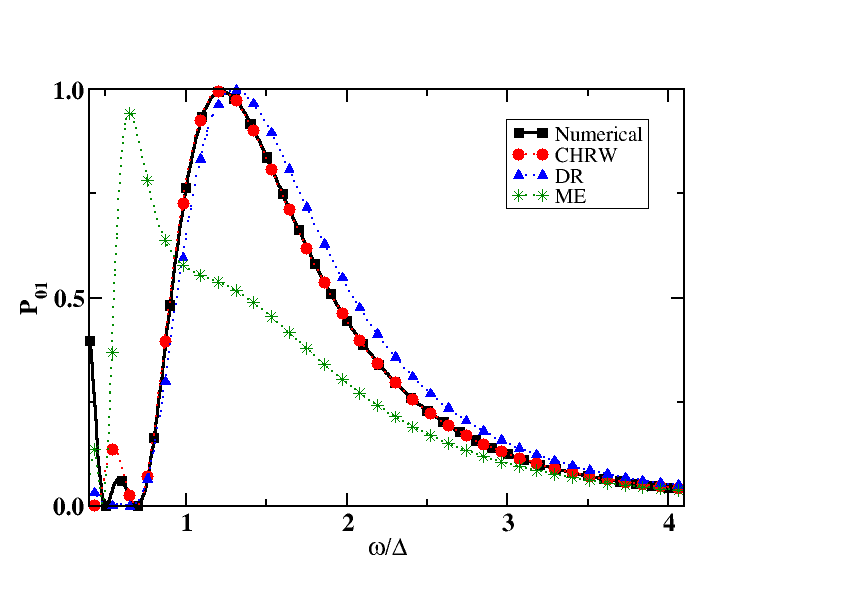}
	\caption{ Plot of the transition probability $P_{01}$ as function of $\omega/\Delta$ for $A=1.16\Delta$ comparing different approximation methods.
 Black squares: numerically exact values. Red circles: counterrotating hybridized rotating wave approximation (CHRW),
 as given by Eq.(\ref{eqP01}). Blue triangles: double rotating frame rotating wave approximation (DR), as given by Eq.(\ref{eqP01ap}). Green stars: Magnus expansion approximation (ME), first order, as given by Eq.(\ref{eqP01MEA}). }
 \label{fig:comp}
\end{figure}
\subsection{Comparison of the different approximations}
We now compare the different approximations for the calculation of the transition probability $P_{01}$.
In Fig.\ref{fig:comp} we show $P_{01}$ as a function of the frequency $\omega$ for $A=1.16\Delta$. We plot the numerically exact values obtained with a highly accurate fourth order Trotter-Suzuki algorithm \cite{Suzuki_2005}.
We find that the first order ME approximation of Eq.(\ref{eqP01MEA}) only agrees with the exact results for $\omega \gtrsim 3\Delta$. On the other hand, the DR approximation of Eq.(\ref{eqP01ap}) agrees reasonably well with the exact dependence with frequency, with errors $\propto 10^{-2}$ in the frequencies of interest and improving accuracy for large frequencies.

The CHRW approximation is very accurate for $\omega\gtrsim \Delta$, and  it is almost indistinguishable from the exact results in the scale of the plot.  From Fig.\ref{fig:fig2} one can see that for this amplitude  and $\omega\gtrsim \Delta$ the error is ${\cal E} < 10^{-5}$.

\bibliography{references}

\providecommand{\noopsort}[1]{}\providecommand{\singleletter}[1]{#1}%
\begin{thebibliography}{68}%
\makeatletter
\providecommand \@ifxundefined [1]{%
 \@ifx{#1\undefined}
}%
\providecommand \@ifnum [1]{%
 \ifnum #1\expandafter \@firstoftwo
 \else \expandafter \@secondoftwo
 \fi
}%
\providecommand \@ifx [1]{%
 \ifx #1\expandafter \@firstoftwo
 \else \expandafter \@secondoftwo
 \fi
}%
\providecommand \natexlab [1]{#1}%
\providecommand \enquote  [1]{``#1''}%
\providecommand \bibnamefont  [1]{#1}%
\providecommand \bibfnamefont [1]{#1}%
\providecommand \citenamefont [1]{#1}%
\providecommand \href@noop [0]{\@secondoftwo}%
\providecommand \href [0]{\begingroup \@sanitize@url \@href}%
\providecommand \@href[1]{\@@startlink{#1}\@@href}%
\providecommand \@@href[1]{\endgroup#1\@@endlink}%
\providecommand \@sanitize@url [0]{\catcode `\\12\catcode `\$12\catcode
  `\&12\catcode `\#12\catcode `\^12\catcode `\_12\catcode `\%12\relax}%
\providecommand \@@startlink[1]{}%
\providecommand \@@endlink[0]{}%
\providecommand \url  [0]{\begingroup\@sanitize@url \@url }%
\providecommand \@url [1]{\endgroup\@href {#1}{\urlprefix }}%
\providecommand \urlprefix  [0]{URL }%
\providecommand \Eprint [0]{\href }%
\providecommand \doibase [0]{http://dx.doi.org/}%
\providecommand \selectlanguage [0]{\@gobble}%
\providecommand \bibinfo  [0]{\@secondoftwo}%
\providecommand \bibfield  [0]{\@secondoftwo}%
\providecommand \translation [1]{[#1]}%
\providecommand \BibitemOpen [0]{}%
\providecommand \bibitemStop [0]{}%
\providecommand \bibitemNoStop [0]{.\EOS\space}%
\providecommand \EOS [0]{\spacefactor3000\relax}%
\providecommand \BibitemShut  [1]{\csname bibitem#1\endcsname}%
\let\auto@bib@innerbib\@empty
\bibitem [{\citenamefont {Krantz}\ \emph {et~al.}(2019)\citenamefont {Krantz},
  \citenamefont {Kjaergaard}, \citenamefont {Yan}, \citenamefont {Orlando},
  \citenamefont {Gustavsson},\ and\ \citenamefont {Oliver}}]{krantz_2019}%
  \BibitemOpen
  \bibfield  {author} {\bibinfo {author} {\bibfnamefont {P.}~\bibnamefont
  {Krantz}}, \bibinfo {author} {\bibfnamefont {M.}~\bibnamefont {Kjaergaard}},
  \bibinfo {author} {\bibfnamefont {F.}~\bibnamefont {Yan}}, \bibinfo {author}
  {\bibfnamefont {T.~P.}\ \bibnamefont {Orlando}}, \bibinfo {author}
  {\bibfnamefont {S.}~\bibnamefont {Gustavsson}}, \ and\ \bibinfo {author}
  {\bibfnamefont {W.~D.}\ \bibnamefont {Oliver}},\ }\href {\doibase
  10.1063/1.5089550} {\bibfield  {journal} {\bibinfo  {journal} {Applied
  Physics Reviews}\ }\textbf {\bibinfo {volume} {6}},\ \bibinfo {pages}
  {021318} (\bibinfo {year} {2019})}\BibitemShut {NoStop}%
\bibitem [{\citenamefont {Kjaergaard}\ \emph {et~al.}(2020)\citenamefont
  {Kjaergaard}, \citenamefont {Schwartz}, \citenamefont {Braum{"u}ller},
  \citenamefont {Krantz}, \citenamefont {Wang}, \citenamefont {Gustavsson},\
  and\ \citenamefont {Oliver}}]{kjaergaard_2020}%
  \BibitemOpen
  \bibfield  {author} {\bibinfo {author} {\bibfnamefont {M.}~\bibnamefont
  {Kjaergaard}}, \bibinfo {author} {\bibfnamefont {M.}~\bibnamefont
  {Schwartz}}, \bibinfo {author} {\bibfnamefont {J.}~\bibnamefont
  {Braum{"u}ller}}, \bibinfo {author} {\bibfnamefont {P.}~\bibnamefont
  {Krantz}}, \bibinfo {author} {\bibfnamefont {J.~I.-J.}\ \bibnamefont {Wang}},
  \bibinfo {author} {\bibfnamefont {S.}~\bibnamefont {Gustavsson}}, \ and\
  \bibinfo {author} {\bibfnamefont {W.}~\bibnamefont {Oliver}},\ }\href@noop {}
  {\bibfield  {journal} {\bibinfo  {journal} {Annual Review of Condensed Matter
  Physics}\ }\textbf {\bibinfo {volume} {11}},\ \bibinfo {pages} {369}
  (\bibinfo {year} {2020})}\BibitemShut {NoStop}%
\bibitem [{\citenamefont {Kwon}\ \emph {et~al.}(2021)\citenamefont {Kwon},
  \citenamefont {Tomonaga}, \citenamefont {Lakshmi~Bhai}, \citenamefont
  {Devitt},\ and\ \citenamefont {Tsai}}]{kwon_2021}%
  \BibitemOpen
  \bibfield  {author} {\bibinfo {author} {\bibfnamefont {S.}~\bibnamefont
  {Kwon}}, \bibinfo {author} {\bibfnamefont {A.}~\bibnamefont {Tomonaga}},
  \bibinfo {author} {\bibfnamefont {G.}~\bibnamefont {Lakshmi~Bhai}}, \bibinfo
  {author} {\bibfnamefont {S.~J.}\ \bibnamefont {Devitt}}, \ and\ \bibinfo
  {author} {\bibfnamefont {J.-S.}\ \bibnamefont {Tsai}},\ }\href {\doibase
  10.1063/5.0029735} {\bibfield  {journal} {\bibinfo  {journal} {Journal of
  Applied Physics}\ }\textbf {\bibinfo {volume} {129}},\ \bibinfo {pages}
  {041102} (\bibinfo {year} {2021})},\ \Eprint
  {http://arxiv.org/abs/https://pubs.aip.org/aip/jap/article-pdf/doi/10.1063/5.0029735/14770083/041102\_1\_online.pdf}
  {https://pubs.aip.org/aip/jap/article-pdf/doi/10.1063/5.0029735/14770083/041102\_1\_online.pdf}
  \BibitemShut {NoStop}%
\bibitem [{\citenamefont {Koch}\ \emph {et~al.}(2007)\citenamefont {Koch},
  \citenamefont {Yu}, \citenamefont {Terri}, \citenamefont {Gambetta},
  \citenamefont {Houck}, \citenamefont {Schuster}, \citenamefont {Majer},
  \citenamefont {Blais}, \citenamefont {Devoret}, \citenamefont {Girvin},\ and\
  \citenamefont {Schoelkopf}}]{Koch_2007}%
  \BibitemOpen
  \bibfield  {author} {\bibinfo {author} {\bibfnamefont {J.}~\bibnamefont
  {Koch}}, \bibinfo {author} {\bibfnamefont {T.}~\bibnamefont {Yu}}, \bibinfo
  {author} {\bibfnamefont {M.}~\bibnamefont {Terri}}, \bibinfo {author}
  {\bibfnamefont {J.}~\bibnamefont {Gambetta}}, \bibinfo {author}
  {\bibfnamefont {A.~A.}\ \bibnamefont {Houck}}, \bibinfo {author}
  {\bibfnamefont {D.~I.}\ \bibnamefont {Schuster}}, \bibinfo {author}
  {\bibfnamefont {J.}~\bibnamefont {Majer}}, \bibinfo {author} {\bibfnamefont
  {A.}~\bibnamefont {Blais}}, \bibinfo {author} {\bibfnamefont
  {M.}~\bibnamefont {Devoret}}, \bibinfo {author} {\bibfnamefont
  {S.}~\bibnamefont {Girvin}}, \ and\ \bibinfo {author} {\bibfnamefont {R.~J.}\
  \bibnamefont {Schoelkopf}},\ }\href {\doibase 10.1103/PhysRevA.76.042319}
  {\bibfield  {journal} {\bibinfo  {journal} {Phys. Rev. A}\ }\textbf {\bibinfo
  {volume} {76}},\ \bibinfo {pages} {042319} (\bibinfo {year}
  {2007})}\BibitemShut {NoStop}%
\bibitem [{\citenamefont {Yan}\ \emph {et~al.}(2016)\citenamefont {Yan},
  \citenamefont {Gustavsson}, \citenamefont {Kamal}, \citenamefont {Birenbaum},
  \citenamefont {Sears}, \citenamefont {Hover}, \citenamefont {Gudmundsen},
  \citenamefont {Rosenberg}, \citenamefont {Samach}, \citenamefont {Weber}
  \emph {et~al.}}]{yan_2016}%
  \BibitemOpen
  \bibfield  {author} {\bibinfo {author} {\bibfnamefont {F.}~\bibnamefont
  {Yan}}, \bibinfo {author} {\bibfnamefont {S.}~\bibnamefont {Gustavsson}},
  \bibinfo {author} {\bibfnamefont {A.}~\bibnamefont {Kamal}}, \bibinfo
  {author} {\bibfnamefont {J.}~\bibnamefont {Birenbaum}}, \bibinfo {author}
  {\bibfnamefont {A.~P.}\ \bibnamefont {Sears}}, \bibinfo {author}
  {\bibfnamefont {D.}~\bibnamefont {Hover}}, \bibinfo {author} {\bibfnamefont
  {T.~J.}\ \bibnamefont {Gudmundsen}}, \bibinfo {author} {\bibfnamefont
  {D.}~\bibnamefont {Rosenberg}}, \bibinfo {author} {\bibfnamefont
  {G.}~\bibnamefont {Samach}}, \bibinfo {author} {\bibfnamefont
  {S.}~\bibnamefont {Weber}},  \emph {et~al.},\ }\href@noop {} {\bibfield
  {journal} {\bibinfo  {journal} {Nature communications}\ }\textbf {\bibinfo
  {volume} {7}},\ \bibinfo {pages} {1} (\bibinfo {year} {2016})}\BibitemShut
  {NoStop}%
\bibitem [{\citenamefont {Manucharyan}\ \emph {et~al.}(2009)\citenamefont
  {Manucharyan}, \citenamefont {Koch}, \citenamefont {Glazman},\ and\
  \citenamefont {Devoret}}]{manucharyan_2009}%
  \BibitemOpen
  \bibfield  {author} {\bibinfo {author} {\bibfnamefont {A.}~\bibnamefont
  {Manucharyan}}, \bibinfo {author} {\bibfnamefont {J.}~\bibnamefont {Koch}},
  \bibinfo {author} {\bibfnamefont {L.~I.}\ \bibnamefont {Glazman}}, \ and\
  \bibinfo {author} {\bibfnamefont {M.~H.}\ \bibnamefont {Devoret}},\ }\href
  {\doibase 10.1126/science.1175552} {\bibfield  {journal} {\bibinfo  {journal}
  {Science}\ }\textbf {\bibinfo {volume} {326}},\ \bibinfo {pages} {113}
  (\bibinfo {year} {2009})}\BibitemShut {NoStop}%
\bibitem [{\citenamefont {Pop}\ \emph {et~al.}(2014)\citenamefont {Pop},
  \citenamefont {Geerlings}, \citenamefont {Catelani}, \citenamefont
  {Schoelkopf}, \citenamefont {Glazman},\ and\ \citenamefont
  {Devoret}}]{pop_2014}%
  \BibitemOpen
  \bibfield  {author} {\bibinfo {author} {\bibfnamefont {I.~M.}\ \bibnamefont
  {Pop}}, \bibinfo {author} {\bibfnamefont {K.}~\bibnamefont {Geerlings}},
  \bibinfo {author} {\bibfnamefont {G.}~\bibnamefont {Catelani}}, \bibinfo
  {author} {\bibfnamefont {R.~J.}\ \bibnamefont {Schoelkopf}}, \bibinfo
  {author} {\bibfnamefont {L.~I.}\ \bibnamefont {Glazman}}, \ and\ \bibinfo
  {author} {\bibfnamefont {M.~H.}\ \bibnamefont {Devoret}},\ }\href@noop {}
  {\bibfield  {journal} {\bibinfo  {journal} {Nature}\ }\textbf {\bibinfo
  {volume} {508}},\ \bibinfo {pages} {369} (\bibinfo {year}
  {2014})}\BibitemShut {NoStop}%
\bibitem [{\citenamefont {Nguyen}\ \emph {et~al.}(2019)\citenamefont {Nguyen},
  \citenamefont {Lin}, \citenamefont {Somoroff}, \citenamefont {Mencia},
  \citenamefont {Grabon},\ and\ \citenamefont {Manucharyan}}]{nguyen_2019}%
  \BibitemOpen
  \bibfield  {author} {\bibinfo {author} {\bibfnamefont {L.~B.}\ \bibnamefont
  {Nguyen}}, \bibinfo {author} {\bibfnamefont {Y.-H.}\ \bibnamefont {Lin}},
  \bibinfo {author} {\bibfnamefont {A.}~\bibnamefont {Somoroff}}, \bibinfo
  {author} {\bibfnamefont {R.}~\bibnamefont {Mencia}}, \bibinfo {author}
  {\bibfnamefont {N.}~\bibnamefont {Grabon}}, \ and\ \bibinfo {author}
  {\bibfnamefont {V.~E.}\ \bibnamefont {Manucharyan}},\ }\href@noop {}
  {\bibfield  {journal} {\bibinfo  {journal} {Physical Review X}\ }\textbf
  {\bibinfo {volume} {9}},\ \bibinfo {pages} {041041} (\bibinfo {year}
  {2019})}\BibitemShut {NoStop}%
\bibitem [{\citenamefont {Bao}\ \emph {et~al.}(2022)\citenamefont {Bao},
  \citenamefont {Deng}, \citenamefont {Ding}, \citenamefont {Gao},
  \citenamefont {Gao}, \citenamefont {Huang}, \citenamefont {Jiang},
  \citenamefont {Ku}, \citenamefont {Li}, \citenamefont {Ma}, \citenamefont
  {Ni}, \citenamefont {Qin}, \citenamefont {Song}, \citenamefont {Sun},
  \citenamefont {Tang}, \citenamefont {Wang}, \citenamefont {Wu}, \citenamefont
  {Xia}, \citenamefont {Yu}, \citenamefont {Zhang}, \citenamefont {Zhang},
  \citenamefont {Zhang}, \citenamefont {Zhou}, \citenamefont {Zhu},
  \citenamefont {Shi}, \citenamefont {Chen}, \citenamefont {Zhao},\ and\
  \citenamefont {Deng}}]{bao_2022}%
  \BibitemOpen
  \bibfield  {author} {\bibinfo {author} {\bibfnamefont {F.}~\bibnamefont
  {Bao}}, \bibinfo {author} {\bibfnamefont {H.}~\bibnamefont {Deng}}, \bibinfo
  {author} {\bibfnamefont {D.}~\bibnamefont {Ding}}, \bibinfo {author}
  {\bibfnamefont {R.}~\bibnamefont {Gao}}, \bibinfo {author} {\bibfnamefont
  {X.}~\bibnamefont {Gao}}, \bibinfo {author} {\bibfnamefont {C.}~\bibnamefont
  {Huang}}, \bibinfo {author} {\bibfnamefont {X.}~\bibnamefont {Jiang}},
  \bibinfo {author} {\bibfnamefont {H.-S.}\ \bibnamefont {Ku}}, \bibinfo
  {author} {\bibfnamefont {Z.}~\bibnamefont {Li}}, \bibinfo {author}
  {\bibfnamefont {X.}~\bibnamefont {Ma}}, \bibinfo {author} {\bibfnamefont
  {X.}~\bibnamefont {Ni}}, \bibinfo {author} {\bibfnamefont {J.}~\bibnamefont
  {Qin}}, \bibinfo {author} {\bibfnamefont {Z.}~\bibnamefont {Song}}, \bibinfo
  {author} {\bibfnamefont {H.}~\bibnamefont {Sun}}, \bibinfo {author}
  {\bibfnamefont {C.}~\bibnamefont {Tang}}, \bibinfo {author} {\bibfnamefont
  {T.}~\bibnamefont {Wang}}, \bibinfo {author} {\bibfnamefont {F.}~\bibnamefont
  {Wu}}, \bibinfo {author} {\bibfnamefont {T.}~\bibnamefont {Xia}}, \bibinfo
  {author} {\bibfnamefont {W.}~\bibnamefont {Yu}}, \bibinfo {author}
  {\bibfnamefont {F.}~\bibnamefont {Zhang}}, \bibinfo {author} {\bibfnamefont
  {G.}~\bibnamefont {Zhang}}, \bibinfo {author} {\bibfnamefont
  {X.}~\bibnamefont {Zhang}}, \bibinfo {author} {\bibfnamefont
  {J.}~\bibnamefont {Zhou}}, \bibinfo {author} {\bibfnamefont {X.}~\bibnamefont
  {Zhu}}, \bibinfo {author} {\bibfnamefont {Y.}~\bibnamefont {Shi}}, \bibinfo
  {author} {\bibfnamefont {J.}~\bibnamefont {Chen}}, \bibinfo {author}
  {\bibfnamefont {H.-H.}\ \bibnamefont {Zhao}}, \ and\ \bibinfo {author}
  {\bibfnamefont {C.}~\bibnamefont {Deng}},\ }\href {\doibase
  10.1103/PhysRevLett.129.010502} {\bibfield  {journal} {\bibinfo  {journal}
  {Phys. Rev. Lett.}\ }\textbf {\bibinfo {volume} {129}},\ \bibinfo {pages}
  {010502} (\bibinfo {year} {2022})}\BibitemShut {NoStop}%
\bibitem [{\citenamefont {Weiss}\ \emph {et~al.}(2022)\citenamefont {Weiss},
  \citenamefont {Zhang}, \citenamefont {Ding}, \citenamefont {Ma},
  \citenamefont {Schuster},\ and\ \citenamefont {Koch}}]{weiss_2022}%
  \BibitemOpen
  \bibfield  {author} {\bibinfo {author} {\bibfnamefont {D.~K.}\ \bibnamefont
  {Weiss}}, \bibinfo {author} {\bibfnamefont {H.}~\bibnamefont {Zhang}},
  \bibinfo {author} {\bibfnamefont {C.}~\bibnamefont {Ding}}, \bibinfo {author}
  {\bibfnamefont {Y.}~\bibnamefont {Ma}}, \bibinfo {author} {\bibfnamefont
  {D.~I.}\ \bibnamefont {Schuster}}, \ and\ \bibinfo {author} {\bibfnamefont
  {J.}~\bibnamefont {Koch}},\ }\href@noop {} {\bibfield  {journal} {\bibinfo
  {journal} {PRX Quantum}\ }\textbf {\bibinfo {volume} {3}},\ \bibinfo {pages}
  {040336} (\bibinfo {year} {2022})}\BibitemShut {NoStop}%
\bibitem [{\citenamefont {Somoroff}\ \emph {et~al.}(2023)\citenamefont
  {Somoroff}, \citenamefont {Ficheux}, \citenamefont {Mencia}, \citenamefont
  {Xiong}, \citenamefont {Kuzmin},\ and\ \citenamefont
  {Manucharyan}}]{somoroff_2023}%
  \BibitemOpen
  \bibfield  {author} {\bibinfo {author} {\bibfnamefont {A.}~\bibnamefont
  {Somoroff}}, \bibinfo {author} {\bibfnamefont {Q.}~\bibnamefont {Ficheux}},
  \bibinfo {author} {\bibfnamefont {R.~A.}\ \bibnamefont {Mencia}}, \bibinfo
  {author} {\bibfnamefont {H.}~\bibnamefont {Xiong}}, \bibinfo {author}
  {\bibfnamefont {R.}~\bibnamefont {Kuzmin}}, \ and\ \bibinfo {author}
  {\bibfnamefont {V.~E.}\ \bibnamefont {Manucharyan}},\ }\href {\doibase
  10.1103/PhysRevLett.130.267001} {\bibfield  {journal} {\bibinfo  {journal}
  {Phys. Rev. Lett.}\ }\textbf {\bibinfo {volume} {130}},\ \bibinfo {pages}
  {267001} (\bibinfo {year} {2023})}\BibitemShut {NoStop}%
\bibitem [{\citenamefont {Leek}\ \emph {et~al.}(2007)\citenamefont {Leek},
  \citenamefont {Fink}, \citenamefont {Blais}, \citenamefont {Bianchetti},
  \citenamefont {G{"o}ppl}, \citenamefont {Gambetta}, \citenamefont {Schuster},
  \citenamefont {Frunzio}, \citenamefont {Schoelkopf},\ and\ \citenamefont
  {Wallraff}}]{leek_2007}%
  \BibitemOpen
  \bibfield  {author} {\bibinfo {author} {\bibfnamefont {P.~J.}\ \bibnamefont
  {Leek}}, \bibinfo {author} {\bibfnamefont {J.~M.}\ \bibnamefont {Fink}},
  \bibinfo {author} {\bibfnamefont {A.}~\bibnamefont {Blais}}, \bibinfo
  {author} {\bibfnamefont {R.}~\bibnamefont {Bianchetti}}, \bibinfo {author}
  {\bibfnamefont {M.}~\bibnamefont {G{"o}ppl}}, \bibinfo {author}
  {\bibfnamefont {J.~M.}\ \bibnamefont {Gambetta}}, \bibinfo {author}
  {\bibfnamefont {D.~I.}\ \bibnamefont {Schuster}}, \bibinfo {author}
  {\bibfnamefont {L.}~\bibnamefont {Frunzio}}, \bibinfo {author} {\bibfnamefont
  {R.~J.}\ \bibnamefont {Schoelkopf}}, \ and\ \bibinfo {author} {\bibfnamefont
  {A.}~\bibnamefont {Wallraff}},\ }\href@noop {} {\bibfield  {journal}
  {\bibinfo  {journal} {Science}\ }\textbf {\bibinfo {volume} {318}},\ \bibinfo
  {pages} {1889} (\bibinfo {year} {2007})}\BibitemShut {NoStop}%
\bibitem [{\citenamefont {Bylander}\ \emph {et~al.}(2011)\citenamefont
  {Bylander}, \citenamefont {Gustavsson}, \citenamefont {Yan}, \citenamefont
  {Yoshihara}, \citenamefont {Harrabi}, \citenamefont {Fitch}, \citenamefont
  {Cory}, \citenamefont {Nakamura}, \citenamefont {Tsai},\ and\ \citenamefont
  {Oliver}}]{bylander_2011}%
  \BibitemOpen
  \bibfield  {author} {\bibinfo {author} {\bibfnamefont {J.}~\bibnamefont
  {Bylander}}, \bibinfo {author} {\bibfnamefont {S.}~\bibnamefont
  {Gustavsson}}, \bibinfo {author} {\bibfnamefont {F.}~\bibnamefont {Yan}},
  \bibinfo {author} {\bibfnamefont {F.}~\bibnamefont {Yoshihara}}, \bibinfo
  {author} {\bibfnamefont {K.}~\bibnamefont {Harrabi}}, \bibinfo {author}
  {\bibfnamefont {G.}~\bibnamefont {Fitch}}, \bibinfo {author} {\bibfnamefont
  {D.~G.}\ \bibnamefont {Cory}}, \bibinfo {author} {\bibfnamefont
  {Y.}~\bibnamefont {Nakamura}}, \bibinfo {author} {\bibfnamefont {J.-S.}\
  \bibnamefont {Tsai}}, \ and\ \bibinfo {author} {\bibfnamefont {W.~D.}\
  \bibnamefont {Oliver}},\ }\href {https://doi.org/10.1038/nphys1994}
  {\bibfield  {journal} {\bibinfo  {journal} {Nature Physics}\ }\textbf
  {\bibinfo {volume} {7}},\ \bibinfo {pages} {565 EP } (\bibinfo {year}
  {2011})}\BibitemShut {NoStop}%
\bibitem [{\citenamefont {Yang}\ \emph {et~al.}(2017)\citenamefont {Yang},
  \citenamefont {Coppersmith},\ and\ \citenamefont {Friesen}}]{yang_2017}%
  \BibitemOpen
  \bibfield  {author} {\bibinfo {author} {\bibfnamefont {Y.-C.}\ \bibnamefont
  {Yang}}, \bibinfo {author} {\bibfnamefont {S.~N.}\ \bibnamefont
  {Coppersmith}}, \ and\ \bibinfo {author} {\bibfnamefont {M.}~\bibnamefont
  {Friesen}},\ }\href {\doibase 10.1103/PhysRevA.95.062321} {\bibfield
  {journal} {\bibinfo  {journal} {Phys. Rev. A}\ }\textbf {\bibinfo {volume}
  {95}},\ \bibinfo {pages} {062321} (\bibinfo {year} {2017})}\BibitemShut
  {NoStop}%
\bibitem [{\citenamefont {Wang}\ \emph {et~al.}(2017)\citenamefont {Wang},
  \citenamefont {Guo}, \citenamefont {Zhang}, \citenamefont {Wang},\ and\
  \citenamefont {Wu}}]{wang_2017}%
  \BibitemOpen
  \bibfield  {author} {\bibinfo {author} {\bibfnamefont {Y.}~\bibnamefont
  {Wang}}, \bibinfo {author} {\bibfnamefont {C.}~\bibnamefont {Guo}}, \bibinfo
  {author} {\bibfnamefont {G.-Q.}\ \bibnamefont {Zhang}}, \bibinfo {author}
  {\bibfnamefont {G.}~\bibnamefont {Wang}}, \ and\ \bibinfo {author}
  {\bibfnamefont {C.}~\bibnamefont {Wu}},\ }\href {\doibase 10.1038/srep44251}
  {\bibfield  {journal} {\bibinfo  {journal} {Scientific Reports}\ }\textbf
  {\bibinfo {volume} {7}},\ \bibinfo {pages} {44251} (\bibinfo {year}
  {2017})}\BibitemShut {NoStop}%
\bibitem [{\citenamefont {Zhu}\ \emph {et~al.}(2021)\citenamefont {Zhu},
  \citenamefont {Jaako}, \citenamefont {He},\ and\ \citenamefont
  {Rabl}}]{zhu_2021}%
  \BibitemOpen
  \bibfield  {author} {\bibinfo {author} {\bibfnamefont {D.}~\bibnamefont
  {Zhu}}, \bibinfo {author} {\bibfnamefont {T.}~\bibnamefont {Jaako}}, \bibinfo
  {author} {\bibfnamefont {Q.}~\bibnamefont {He}}, \ and\ \bibinfo {author}
  {\bibfnamefont {P.}~\bibnamefont {Rabl}},\ }\href {\doibase
  10.1103/PhysRevApplied.16.014024} {\bibfield  {journal} {\bibinfo  {journal}
  {Phys. Rev. Appl.}\ }\textbf {\bibinfo {volume} {16}},\ \bibinfo {pages}
  {014024} (\bibinfo {year} {2021})}\BibitemShut {NoStop}%
\bibitem [{\citenamefont {Shen}\ \emph {et~al.}(2021)\citenamefont {Shen},
  \citenamefont {Chen},\ and\ \citenamefont {Xue}}]{shen_2021}%
  \BibitemOpen
  \bibfield  {author} {\bibinfo {author} {\bibfnamefont {P.}~\bibnamefont
  {Shen}}, \bibinfo {author} {\bibfnamefont {T.}~\bibnamefont {Chen}}, \ and\
  \bibinfo {author} {\bibfnamefont {Z.-Y.}\ \bibnamefont {Xue}},\ }\href
  {\doibase 10.1103/PhysRevApplied.16.044004} {\bibfield  {journal} {\bibinfo
  {journal} {Phys. Rev. Appl.}\ }\textbf {\bibinfo {volume} {16}},\ \bibinfo
  {pages} {044004} (\bibinfo {year} {2021})}\BibitemShut {NoStop}%
\bibitem [{\citenamefont {Ficheux}\ \emph {et~al.}(2021)\citenamefont
  {Ficheux}, \citenamefont {Nguyen}, \citenamefont {Somoroff}, \citenamefont
  {Xiong}, \citenamefont {Nesterov}, \citenamefont {Vavilov},\ and\
  \citenamefont {Manucharyan}}]{ficheux_2021}%
  \BibitemOpen
  \bibfield  {author} {\bibinfo {author} {\bibfnamefont {Q.}~\bibnamefont
  {Ficheux}}, \bibinfo {author} {\bibfnamefont {L.~B.}\ \bibnamefont {Nguyen}},
  \bibinfo {author} {\bibfnamefont {A.}~\bibnamefont {Somoroff}}, \bibinfo
  {author} {\bibfnamefont {H.}~\bibnamefont {Xiong}}, \bibinfo {author}
  {\bibfnamefont {K.~N.}\ \bibnamefont {Nesterov}}, \bibinfo {author}
  {\bibfnamefont {M.~G.}\ \bibnamefont {Vavilov}}, \ and\ \bibinfo {author}
  {\bibfnamefont {V.~E.}\ \bibnamefont {Manucharyan}},\ }\href {\doibase
  10.1103/PhysRevX.11.021026} {\bibfield  {journal} {\bibinfo  {journal} {Phys.
  Rev. X}\ }\textbf {\bibinfo {volume} {11}},\ \bibinfo {pages} {021026}
  (\bibinfo {year} {2021})}\BibitemShut {NoStop}%
\bibitem [{\citenamefont {Bastrakova}\ \emph {et~al.}(2022)\citenamefont
  {Bastrakova}, \citenamefont {Klenov}, \citenamefont {Ruzhickiy},
  \citenamefont {Soloviev},\ and\ \citenamefont {Satanin}}]{bastrakova_2022}%
  \BibitemOpen
  \bibfield  {author} {\bibinfo {author} {\bibfnamefont {M.}~\bibnamefont
  {Bastrakova}}, \bibinfo {author} {\bibfnamefont {N.}~\bibnamefont {Klenov}},
  \bibinfo {author} {\bibfnamefont {V.}~\bibnamefont {Ruzhickiy}}, \bibinfo
  {author} {\bibfnamefont {I.}~\bibnamefont {Soloviev}}, \ and\ \bibinfo
  {author} {\bibfnamefont {A.}~\bibnamefont {Satanin}},\ }\href {\doibase
  10.1088/1361-6668/ac5505} {\bibfield  {journal} {\bibinfo  {journal}
  {Superconductor Science and Technology}\ }\textbf {\bibinfo {volume} {35}},\
  \bibinfo {pages} {055003} (\bibinfo {year} {2022})}\BibitemShut {NoStop}%
\bibitem [{\citenamefont {Chen}\ \emph {et~al.}(2022)\citenamefont {Chen},
  \citenamefont {Miranowicz}, \citenamefont {Chen}, \citenamefont {Xia},\ and\
  \citenamefont {Nori}}]{chen_2022}%
  \BibitemOpen
  \bibfield  {author} {\bibinfo {author} {\bibfnamefont {Y.-H.}\ \bibnamefont
  {Chen}}, \bibinfo {author} {\bibfnamefont {A.}~\bibnamefont {Miranowicz}},
  \bibinfo {author} {\bibfnamefont {X.}~\bibnamefont {Chen}}, \bibinfo {author}
  {\bibfnamefont {Y.}~\bibnamefont {Xia}}, \ and\ \bibinfo {author}
  {\bibfnamefont {F.}~\bibnamefont {Nori}},\ }\href {\doibase
  10.1103/PhysRevApplied.18.064059} {\bibfield  {journal} {\bibinfo  {journal}
  {Phys. Rev. Appl.}\ }\textbf {\bibinfo {volume} {18}},\ \bibinfo {pages}
  {064059} (\bibinfo {year} {2022})}\BibitemShut {NoStop}%
\bibitem [{\citenamefont {Bloch}\ and\ \citenamefont
  {Siegert}(1940)}]{bloch_1940}%
  \BibitemOpen
  \bibfield  {author} {\bibinfo {author} {\bibfnamefont {F.}~\bibnamefont
  {Bloch}}\ and\ \bibinfo {author} {\bibfnamefont {A.}~\bibnamefont
  {Siegert}},\ }\href@noop {} {\bibfield  {journal} {\bibinfo  {journal}
  {Physical Review}\ }\textbf {\bibinfo {volume} {57}},\ \bibinfo {pages} {522}
  (\bibinfo {year} {1940})}\BibitemShut {NoStop}%
\bibitem [{\citenamefont {Avinadav}\ \emph {et~al.}(2014)\citenamefont
  {Avinadav}, \citenamefont {Fischer}, \citenamefont {London},\ and\
  \citenamefont {Gershoni}}]{avi_2014}%
  \BibitemOpen
  \bibfield  {author} {\bibinfo {author} {\bibfnamefont {C.}~\bibnamefont
  {Avinadav}}, \bibinfo {author} {\bibfnamefont {R.}~\bibnamefont {Fischer}},
  \bibinfo {author} {\bibfnamefont {P.}~\bibnamefont {London}}, \ and\ \bibinfo
  {author} {\bibfnamefont {D.}~\bibnamefont {Gershoni}},\ }\href {\doibase
  10.1103/PhysRevB.89.245311} {\bibfield  {journal} {\bibinfo  {journal} {Phys.
  Rev. B}\ }\textbf {\bibinfo {volume} {89}},\ \bibinfo {pages} {245311}
  (\bibinfo {year} {2014})}\BibitemShut {NoStop}%
\bibitem [{\citenamefont {Campbell}\ \emph {et~al.}(2020)\citenamefont
  {Campbell}, \citenamefont {Shim}, \citenamefont {Kannan}, \citenamefont
  {Winik}, \citenamefont {Kim}, \citenamefont {Melville}, \citenamefont
  {Niedzielski}, \citenamefont {Yoder}, \citenamefont {Tahan}, \citenamefont
  {Gustavsson},\ and\ \citenamefont {Oliver}}]{campbell_2020}%
  \BibitemOpen
  \bibfield  {author} {\bibinfo {author} {\bibfnamefont {D.~L.}\ \bibnamefont
  {Campbell}}, \bibinfo {author} {\bibfnamefont {Y.-P.}\ \bibnamefont {Shim}},
  \bibinfo {author} {\bibfnamefont {B.}~\bibnamefont {Kannan}}, \bibinfo
  {author} {\bibfnamefont {R.}~\bibnamefont {Winik}}, \bibinfo {author}
  {\bibfnamefont {D.~K.}\ \bibnamefont {Kim}}, \bibinfo {author} {\bibfnamefont
  {A.}~\bibnamefont {Melville}}, \bibinfo {author} {\bibfnamefont {B.~M.}\
  \bibnamefont {Niedzielski}}, \bibinfo {author} {\bibfnamefont {J.~L.}\
  \bibnamefont {Yoder}}, \bibinfo {author} {\bibfnamefont {C.}~\bibnamefont
  {Tahan}}, \bibinfo {author} {\bibfnamefont {S.}~\bibnamefont {Gustavsson}}, \
  and\ \bibinfo {author} {\bibfnamefont {W.~D.}\ \bibnamefont {Oliver}},\
  }\href {\doibase 10.1103/PhysRevX.10.041051} {\bibfield  {journal} {\bibinfo
  {journal} {Phys. Rev. X}\ }\textbf {\bibinfo {volume} {10}},\ \bibinfo
  {pages} {041051} (\bibinfo {year} {2020})}\BibitemShut {NoStop}%
\bibitem [{\citenamefont {Zhang}\ \emph {et~al.}(2021)\citenamefont {Zhang},
  \citenamefont {Ma}, \citenamefont {Weiss}, \citenamefont {Ding},
  \citenamefont {Li}, \citenamefont {Huang}, \citenamefont {Schuster},
  \citenamefont {Jiang},\ and\ \citenamefont {Koch}}]{zhang_2021}%
  \BibitemOpen
  \bibfield  {author} {\bibinfo {author} {\bibfnamefont {H.}~\bibnamefont
  {Zhang}}, \bibinfo {author} {\bibfnamefont {Y.}~\bibnamefont {Ma}}, \bibinfo
  {author} {\bibfnamefont {D.~K.}\ \bibnamefont {Weiss}}, \bibinfo {author}
  {\bibfnamefont {C.}~\bibnamefont {Ding}}, \bibinfo {author} {\bibfnamefont
  {Y.}~\bibnamefont {Li}}, \bibinfo {author} {\bibfnamefont {W.}~\bibnamefont
  {Huang}}, \bibinfo {author} {\bibfnamefont {D.~I.}\ \bibnamefont {Schuster}},
  \bibinfo {author} {\bibfnamefont {L.}~\bibnamefont {Jiang}}, \ and\ \bibinfo
  {author} {\bibfnamefont {J.}~\bibnamefont {Koch}},\ }\href {\doibase
  10.1103/PhysRevX.11.011010} {\bibfield  {journal} {\bibinfo  {journal}
  {Physical Review X}\ }\textbf {\bibinfo {volume} {11}},\ \bibinfo {pages}
  {011010} (\bibinfo {year} {2021})}\BibitemShut {NoStop}%
\bibitem [{\citenamefont {Petrescu}\ \emph {et~al.}(2023)\citenamefont
  {Petrescu}, \citenamefont {Le~Calonnec}, \citenamefont {Leroux},
  \citenamefont {Di~Paolo}, \citenamefont {Mundada}, \citenamefont {Sussman},
  \citenamefont {Vrajitoarea}, \citenamefont {Houck},\ and\ \citenamefont
  {Blais}}]{petrescu_2023}%
  \BibitemOpen
  \bibfield  {author} {\bibinfo {author} {\bibfnamefont {A.}~\bibnamefont
  {Petrescu}}, \bibinfo {author} {\bibfnamefont {C.}~\bibnamefont
  {Le~Calonnec}}, \bibinfo {author} {\bibfnamefont {C.}~\bibnamefont {Leroux}},
  \bibinfo {author} {\bibfnamefont {A.}~\bibnamefont {Di~Paolo}}, \bibinfo
  {author} {\bibfnamefont {P.}~\bibnamefont {Mundada}}, \bibinfo {author}
  {\bibfnamefont {S.}~\bibnamefont {Sussman}}, \bibinfo {author} {\bibfnamefont
  {A.}~\bibnamefont {Vrajitoarea}}, \bibinfo {author} {\bibfnamefont {A.~A.}\
  \bibnamefont {Houck}}, \ and\ \bibinfo {author} {\bibfnamefont
  {A.}~\bibnamefont {Blais}},\ }\href {\doibase
  10.1103/PhysRevApplied.19.044003} {\bibfield  {journal} {\bibinfo  {journal}
  {Phys. Rev. Appl.}\ }\textbf {\bibinfo {volume} {19}},\ \bibinfo {pages}
  {044003} (\bibinfo {year} {2023})}\BibitemShut {NoStop}%
\bibitem [{\citenamefont {Oliver}\ \emph {et~al.}(2005)\citenamefont {Oliver},
  \citenamefont {Yu}, \citenamefont {Lee}, \citenamefont {Berggren},
  \citenamefont {Levitov},\ and\ \citenamefont {Orlando}}]{Oliver_2005}%
  \BibitemOpen
  \bibfield  {author} {\bibinfo {author} {\bibfnamefont {W.~D.}\ \bibnamefont
  {Oliver}}, \bibinfo {author} {\bibfnamefont {Y.}~\bibnamefont {Yu}}, \bibinfo
  {author} {\bibfnamefont {J.~C.}\ \bibnamefont {Lee}}, \bibinfo {author}
  {\bibfnamefont {K.~K.}\ \bibnamefont {Berggren}}, \bibinfo {author}
  {\bibfnamefont {L.~S.}\ \bibnamefont {Levitov}}, \ and\ \bibinfo {author}
  {\bibfnamefont {T.~P.}\ \bibnamefont {Orlando}},\ }\href@noop {} {\bibfield
  {journal} {\bibinfo  {journal} {Science}\ }\textbf {\bibinfo {volume}
  {310}},\ \bibinfo {pages} {1653} (\bibinfo {year} {2005})}\BibitemShut
  {NoStop}%
\bibitem [{\citenamefont {Sillanp\"a\"a}\ \emph {et~al.}(2006)\citenamefont
  {Sillanp\"a\"a}, \citenamefont {Lehtinen}, \citenamefont {Paila},
  \citenamefont {Makhlin},\ and\ \citenamefont {Hakonen}}]{Sillanpaa_2006}%
  \BibitemOpen
  \bibfield  {author} {\bibinfo {author} {\bibfnamefont {M.}~\bibnamefont
  {Sillanp\"a\"a}}, \bibinfo {author} {\bibfnamefont {T.}~\bibnamefont
  {Lehtinen}}, \bibinfo {author} {\bibfnamefont {A.}~\bibnamefont {Paila}},
  \bibinfo {author} {\bibfnamefont {Y.}~\bibnamefont {Makhlin}}, \ and\
  \bibinfo {author} {\bibfnamefont {P.}~\bibnamefont {Hakonen}},\ }\href
  {\doibase 10.1103/PhysRevLett.96.187002} {\bibfield  {journal} {\bibinfo
  {journal} {Phys. Rev. Lett.}\ }\textbf {\bibinfo {volume} {96}},\ \bibinfo
  {pages} {187002} (\bibinfo {year} {2006})}\BibitemShut {NoStop}%
\bibitem [{\citenamefont {Ferr\'on}\ \emph {et~al.}(2012)\citenamefont
  {Ferr\'on}, \citenamefont {Dom\'{\i}nguez},\ and\ \citenamefont
  {S\'anchez}}]{Ferron_2012}%
  \BibitemOpen
  \bibfield  {author} {\bibinfo {author} {\bibfnamefont {A.}~\bibnamefont
  {Ferr\'on}}, \bibinfo {author} {\bibfnamefont {D.}~\bibnamefont
  {Dom\'{\i}nguez}}, \ and\ \bibinfo {author} {\bibfnamefont {M.~J.}\
  \bibnamefont {S\'anchez}},\ }\href {\doibase 10.1103/PhysRevLett.109.237005}
  {\bibfield  {journal} {\bibinfo  {journal} {Phys. Rev. Lett.}\ }\textbf
  {\bibinfo {volume} {109}},\ \bibinfo {pages} {237005} (\bibinfo {year}
  {2012})}\BibitemShut {NoStop}%
\bibitem [{\citenamefont {Shevchenko}\ \emph {et~al.}(2012)\citenamefont
  {Shevchenko}, \citenamefont {Omelyanchouk},\ and\ \citenamefont
  {Il’ichev}}]{shevchenko_2012}%
  \BibitemOpen
  \bibfield  {author} {\bibinfo {author} {\bibfnamefont {S.~N.}\ \bibnamefont
  {Shevchenko}}, \bibinfo {author} {\bibfnamefont {A.~N.}\ \bibnamefont
  {Omelyanchouk}}, \ and\ \bibinfo {author} {\bibfnamefont {E.}~\bibnamefont
  {Il’ichev}},\ }\href {\doibase 10.1063/1.3701717} {\bibfield  {journal}
  {\bibinfo  {journal} {Low Temperature Physics}\ }\textbf {\bibinfo {volume}
  {38}},\ \bibinfo {pages} {283} (\bibinfo {year} {2012})},\ \Eprint
  {http://arxiv.org/abs/https://doi.org/10.1063/1.3701717}
  {https://doi.org/10.1063/1.3701717} \BibitemShut {NoStop}%
\bibitem [{\citenamefont {Ivakhnenko}\ \emph {et~al.}(2023)\citenamefont
  {Ivakhnenko}, \citenamefont {Shevchenko},\ and\ \citenamefont
  {Nori}}]{ivakhnenko_2023}%
  \BibitemOpen
  \bibfield  {author} {\bibinfo {author} {\bibfnamefont {O.~V.}\ \bibnamefont
  {Ivakhnenko}}, \bibinfo {author} {\bibfnamefont {S.~N.}\ \bibnamefont
  {Shevchenko}}, \ and\ \bibinfo {author} {\bibfnamefont {F.}~\bibnamefont
  {Nori}},\ }\href {\doibase https://doi.org/10.1016/j.physrep.2022.10.002}
  {\bibfield  {journal} {\bibinfo  {journal} {Physics Reports}\ }\textbf
  {\bibinfo {volume} {995}},\ \bibinfo {pages} {1} (\bibinfo {year} {2023})},\
  \bibinfo {note} {nonadiabatic Landau-Zener-Stückelberg-Majorana transitions,
  dynamics, and interference}\BibitemShut {NoStop}%
\bibitem [{\citenamefont {Oliver}\ and\ \citenamefont
  {Valenzuela}(2009)}]{Oliver_2009}%
  \BibitemOpen
  \bibfield  {author} {\bibinfo {author} {\bibfnamefont {W.~D.}\ \bibnamefont
  {Oliver}}\ and\ \bibinfo {author} {\bibfnamefont {S.~O.}\ \bibnamefont
  {Valenzuela}},\ }\href@noop {} {\bibfield  {journal} {\bibinfo  {journal}
  {Quantum Information Processing}\ }\textbf {\bibinfo {volume} {8}},\ \bibinfo
  {pages} {261} (\bibinfo {year} {2009})}\BibitemShut {NoStop}%
\bibitem [{\citenamefont {Berns}\ \emph {et~al.}(2008)\citenamefont {Berns},
  \citenamefont {Rudner}, \citenamefont {Valenzuela}, \citenamefont {Berggren},
  \citenamefont {Oliver}, \citenamefont {Levitov},\ and\ \citenamefont
  {Orlando}}]{Berns_2008}%
  \BibitemOpen
  \bibfield  {author} {\bibinfo {author} {\bibfnamefont {D.~M.}\ \bibnamefont
  {Berns}}, \bibinfo {author} {\bibfnamefont {M.~S.}\ \bibnamefont {Rudner}},
  \bibinfo {author} {\bibfnamefont {S.~O.}\ \bibnamefont {Valenzuela}},
  \bibinfo {author} {\bibfnamefont {K.~K.}\ \bibnamefont {Berggren}}, \bibinfo
  {author} {\bibfnamefont {W.~D.}\ \bibnamefont {Oliver}}, \bibinfo {author}
  {\bibfnamefont {L.~S.}\ \bibnamefont {Levitov}}, \ and\ \bibinfo {author}
  {\bibfnamefont {T.~P.}\ \bibnamefont {Orlando}},\ }\href@noop {} {\bibfield
  {journal} {\bibinfo  {journal} {Nature}\ }\textbf {\bibinfo {volume} {455}},\
  \bibinfo {pages} {51} (\bibinfo {year} {2008})}\BibitemShut {NoStop}%
\bibitem [{\citenamefont {Bylander}\ \emph {et~al.}(2009)\citenamefont
  {Bylander}, \citenamefont {Rudner}, \citenamefont {Shytov}, \citenamefont
  {Valenzuela}, \citenamefont {Berns}, \citenamefont {Berggren}, \citenamefont
  {Levitov},\ and\ \citenamefont {Oliver}}]{bylander_2009}%
  \BibitemOpen
  \bibfield  {author} {\bibinfo {author} {\bibfnamefont {J.}~\bibnamefont
  {Bylander}}, \bibinfo {author} {\bibfnamefont {M.~S.}\ \bibnamefont
  {Rudner}}, \bibinfo {author} {\bibfnamefont {A.}~\bibnamefont {Shytov}},
  \bibinfo {author} {\bibfnamefont {S.~O.}\ \bibnamefont {Valenzuela}},
  \bibinfo {author} {\bibfnamefont {D.}~\bibnamefont {Berns}}, \bibinfo
  {author} {\bibfnamefont {K.}~\bibnamefont {Berggren}}, \bibinfo {author}
  {\bibfnamefont {L.}~\bibnamefont {Levitov}}, \ and\ \bibinfo {author}
  {\bibfnamefont {W.}~\bibnamefont {Oliver}},\ }\href@noop {} {\bibfield
  {journal} {\bibinfo  {journal} {Physical Review B}\ }\textbf {\bibinfo
  {volume} {80}},\ \bibinfo {pages} {220506} (\bibinfo {year}
  {2009})}\BibitemShut {NoStop}%
\bibitem [{\citenamefont {Gustavsson}\ \emph {et~al.}(2013)\citenamefont
  {Gustavsson}, \citenamefont {Bylander},\ and\ \citenamefont
  {Oliver}}]{gustavsson_2013}%
  \BibitemOpen
  \bibfield  {author} {\bibinfo {author} {\bibfnamefont {S.}~\bibnamefont
  {Gustavsson}}, \bibinfo {author} {\bibfnamefont {J.}~\bibnamefont
  {Bylander}}, \ and\ \bibinfo {author} {\bibfnamefont {W.~D.}\ \bibnamefont
  {Oliver}},\ }\href@noop {} {\bibfield  {journal} {\bibinfo  {journal}
  {Physical Review Letters}\ }\textbf {\bibinfo {volume} {110}},\ \bibinfo
  {pages} {017003} (\bibinfo {year} {2013})}\BibitemShut {NoStop}%
\bibitem [{\citenamefont {Gramajo}\ \emph {et~al.}(2020)\citenamefont
  {Gramajo}, \citenamefont {Campbell}, \citenamefont {Kannan}, \citenamefont
  {Kim}, \citenamefont {Melville}, \citenamefont {Niedzielski}, \citenamefont
  {Yoder}, \citenamefont {S\'anchez}, \citenamefont {Dom\'{\i}nguez},
  \citenamefont {Gustavsson},\ and\ \citenamefont {Oliver}}]{gramajo_2020}%
  \BibitemOpen
  \bibfield  {author} {\bibinfo {author} {\bibfnamefont {A.~L.}\ \bibnamefont
  {Gramajo}}, \bibinfo {author} {\bibfnamefont {D.}~\bibnamefont {Campbell}},
  \bibinfo {author} {\bibfnamefont {B.}~\bibnamefont {Kannan}}, \bibinfo
  {author} {\bibfnamefont {D.~K.}\ \bibnamefont {Kim}}, \bibinfo {author}
  {\bibfnamefont {A.}~\bibnamefont {Melville}}, \bibinfo {author}
  {\bibfnamefont {B.~M.}\ \bibnamefont {Niedzielski}}, \bibinfo {author}
  {\bibfnamefont {J.~L.}\ \bibnamefont {Yoder}}, \bibinfo {author}
  {\bibfnamefont {M.~J.}\ \bibnamefont {S\'anchez}}, \bibinfo {author}
  {\bibfnamefont {D.}~\bibnamefont {Dom\'{\i}nguez}}, \bibinfo {author}
  {\bibfnamefont {S.}~\bibnamefont {Gustavsson}}, \ and\ \bibinfo {author}
  {\bibfnamefont {W.~D.}\ \bibnamefont {Oliver}},\ }\href {\doibase
  10.1103/PhysRevApplied.14.014047} {\bibfield  {journal} {\bibinfo  {journal}
  {Phys. Rev. Applied}\ }\textbf {\bibinfo {volume} {14}},\ \bibinfo {pages}
  {014047} (\bibinfo {year} {2020})}\BibitemShut {NoStop}%
\bibitem [{\citenamefont {Shirley}(1965)}]{Shirley_1965}%
  \BibitemOpen
  \bibfield  {author} {\bibinfo {author} {\bibfnamefont {J.~H.}\ \bibnamefont
  {Shirley}},\ }\href@noop {} {\bibfield  {journal} {\bibinfo  {journal}
  {Physical Review}\ }\textbf {\bibinfo {volume} {138}},\ \bibinfo {pages}
  {B979} (\bibinfo {year} {1965})}\BibitemShut {NoStop}%
\bibitem [{\citenamefont {Son}\ \emph {et~al.}(2009)\citenamefont {Son},
  \citenamefont {Han},\ and\ \citenamefont {Chu}}]{son_2009}%
  \BibitemOpen
  \bibfield  {author} {\bibinfo {author} {\bibfnamefont {S.-K.}\ \bibnamefont
  {Son}}, \bibinfo {author} {\bibfnamefont {S.}~\bibnamefont {Han}}, \ and\
  \bibinfo {author} {\bibfnamefont {S.-I.}\ \bibnamefont {Chu}},\ }\href
  {\doibase 10.1103/PhysRevA.79.032301} {\bibfield  {journal} {\bibinfo
  {journal} {Phys. Rev. A}\ }\textbf {\bibinfo {volume} {79}},\ \bibinfo
  {pages} {032301} (\bibinfo {year} {2009})}\BibitemShut {NoStop}%
\bibitem [{\citenamefont {Ferr\'on}\ \emph {et~al.}(2016)\citenamefont
  {Ferr\'on}, \citenamefont {Dom\'{\i}nguez},\ and\ \citenamefont
  {S\'anchez}}]{Ferron_2016}%
  \BibitemOpen
  \bibfield  {author} {\bibinfo {author} {\bibfnamefont {A.}~\bibnamefont
  {Ferr\'on}}, \bibinfo {author} {\bibfnamefont {D.}~\bibnamefont
  {Dom\'{\i}nguez}}, \ and\ \bibinfo {author} {\bibfnamefont {M.~J.}\
  \bibnamefont {S\'anchez}},\ }\href {\doibase 10.1103/PhysRevB.93.064521}
  {\bibfield  {journal} {\bibinfo  {journal} {Phys. Rev. B}\ }\textbf {\bibinfo
  {volume} {93}},\ \bibinfo {pages} {064521} (\bibinfo {year}
  {2016})}\BibitemShut {NoStop}%
\bibitem [{\citenamefont {Yan}\ \emph {et~al.}(2015)\citenamefont {Yan},
  \citenamefont {Lu},\ and\ \citenamefont {Zheng}}]{yan_2015}%
  \BibitemOpen
  \bibfield  {author} {\bibinfo {author} {\bibfnamefont {Y.}~\bibnamefont
  {Yan}}, \bibinfo {author} {\bibfnamefont {Z.}~\bibnamefont {Lu}}, \ and\
  \bibinfo {author} {\bibfnamefont {H.}~\bibnamefont {Zheng}},\ }\href@noop {}
  {\bibfield  {journal} {\bibinfo  {journal} {Physical Review A}\ }\textbf
  {\bibinfo {volume} {91}},\ \bibinfo {pages} {053834} (\bibinfo {year}
  {2015})}\BibitemShut {NoStop}%
\bibitem [{\citenamefont {L\"u}\ and\ \citenamefont {Zheng}(2012)}]{lu_2012}%
  \BibitemOpen
  \bibfield  {author} {\bibinfo {author} {\bibfnamefont {Z.}~\bibnamefont
  {L\"u}}\ and\ \bibinfo {author} {\bibfnamefont {H.}~\bibnamefont {Zheng}},\
  }\href {\doibase 10.1103/PhysRevA.86.023831} {\bibfield  {journal} {\bibinfo
  {journal} {Phys. Rev. A}\ }\textbf {\bibinfo {volume} {86}},\ \bibinfo
  {pages} {023831} (\bibinfo {year} {2012})}\BibitemShut {NoStop}%
\bibitem [{\citenamefont {Hatano}\ and\ \citenamefont
  {Suzuki}(2005{\natexlab{a}})}]{hatano2005}%
  \BibitemOpen
  \bibfield  {author} {\bibinfo {author} {\bibfnamefont {N.}~\bibnamefont
  {Hatano}}\ and\ \bibinfo {author} {\bibfnamefont {M.}~\bibnamefont
  {Suzuki}},\ }\enquote {\bibinfo {title} {Finding exponential product formulas
  of higher orders},}\ in\ \href {\doibase 10.1007/11526216_2} {\emph {\bibinfo
  {booktitle} {Quantum Annealing and Other Optimization Methods}}},\ Vol.\
  \bibinfo {volume} {679},\ \bibinfo {editor} {edited by\ \bibinfo {editor}
  {\bibfnamefont {A.}~\bibnamefont {Das}}\ and\ \bibinfo {editor}
  {\bibfnamefont {B.}~\bibnamefont {K.~Chakrabarti}}}\ (\bibinfo  {publisher}
  {Springer Berlin Heidelberg},\ \bibinfo {address} {Berlin, Heidelberg},\
  \bibinfo {year} {2005})\ pp.\ \bibinfo {pages} {37--68}\BibitemShut {NoStop}%
\bibitem [{\citenamefont {Pedersen}\ \emph {et~al.}(2007)\citenamefont
  {Pedersen}, \citenamefont {Møller},\ and\ \citenamefont
  {Mølmer}}]{pedersen_2007}%
  \BibitemOpen
  \bibfield  {author} {\bibinfo {author} {\bibfnamefont {L.~H.}\ \bibnamefont
  {Pedersen}}, \bibinfo {author} {\bibfnamefont {N.~M.}\ \bibnamefont
  {Møller}}, \ and\ \bibinfo {author} {\bibfnamefont {K.}~\bibnamefont
  {Mølmer}},\ }\href {\doibase https://doi.org/10.1016/j.physleta.2007.02.069}
  {\bibfield  {journal} {\bibinfo  {journal} {Physics Letters A}\ }\textbf
  {\bibinfo {volume} {367}},\ \bibinfo {pages} {47} (\bibinfo {year}
  {2007})}\BibitemShut {NoStop}%
\bibitem [{\citenamefont {Deng}\ \emph {et~al.}(2015)\citenamefont {Deng},
  \citenamefont {Orgiazzi}, \citenamefont {Shen}, \citenamefont {Ashhab},\ and\
  \citenamefont {Lupascu}}]{deng_2015}%
  \BibitemOpen
  \bibfield  {author} {\bibinfo {author} {\bibfnamefont {C.}~\bibnamefont
  {Deng}}, \bibinfo {author} {\bibfnamefont {J.-L.}\ \bibnamefont {Orgiazzi}},
  \bibinfo {author} {\bibfnamefont {F.}~\bibnamefont {Shen}}, \bibinfo {author}
  {\bibfnamefont {S.}~\bibnamefont {Ashhab}}, \ and\ \bibinfo {author}
  {\bibfnamefont {A.}~\bibnamefont {Lupascu}},\ }\href {\doibase
  10.1103/PhysRevLett.115.133601} {\bibfield  {journal} {\bibinfo  {journal}
  {Phys. Rev. Lett.}\ }\textbf {\bibinfo {volume} {115}},\ \bibinfo {pages}
  {133601} (\bibinfo {year} {2015})}\BibitemShut {NoStop}%
\bibitem [{\citenamefont {Deng}\ \emph {et~al.}(2016)\citenamefont {Deng},
  \citenamefont {Shen}, \citenamefont {Ashhab},\ and\ \citenamefont
  {Lupascu}}]{deng_2016}%
  \BibitemOpen
  \bibfield  {author} {\bibinfo {author} {\bibfnamefont {C.}~\bibnamefont
  {Deng}}, \bibinfo {author} {\bibfnamefont {F.}~\bibnamefont {Shen}}, \bibinfo
  {author} {\bibfnamefont {S.}~\bibnamefont {Ashhab}}, \ and\ \bibinfo {author}
  {\bibfnamefont {A.}~\bibnamefont {Lupascu}},\ }\href {\doibase
  10.1103/PhysRevA.94.032323} {\bibfield  {journal} {\bibinfo  {journal} {Phys.
  Rev. A}\ }\textbf {\bibinfo {volume} {94}},\ \bibinfo {pages} {032323}
  (\bibinfo {year} {2016})}\BibitemShut {NoStop}%
\bibitem [{\citenamefont {McKay}\ \emph {et~al.}(2017)\citenamefont {McKay},
  \citenamefont {Wood}, \citenamefont {Sheldon}, \citenamefont {Chow},\ and\
  \citenamefont {Gambetta}}]{mckay_2017}%
  \BibitemOpen
  \bibfield  {author} {\bibinfo {author} {\bibfnamefont {D.~C.}\ \bibnamefont
  {McKay}}, \bibinfo {author} {\bibfnamefont {C.~J.}\ \bibnamefont {Wood}},
  \bibinfo {author} {\bibfnamefont {S.}~\bibnamefont {Sheldon}}, \bibinfo
  {author} {\bibfnamefont {J.~M.}\ \bibnamefont {Chow}}, \ and\ \bibinfo
  {author} {\bibfnamefont {J.~M.}\ \bibnamefont {Gambetta}},\ }\href {\doibase
  10.1103/PhysRevA.96.022330} {\bibfield  {journal} {\bibinfo  {journal} {Phys.
  Rev. A}\ }\textbf {\bibinfo {volume} {96}},\ \bibinfo {pages} {022330}
  (\bibinfo {year} {2017})}\BibitemShut {NoStop}%
\bibitem [{\citenamefont {Huang}\ \emph {et~al.}(2023)\citenamefont {Huang},
  \citenamefont {Wang}, \citenamefont {Wu}, \citenamefont {Ding}, \citenamefont
  {Ye}, \citenamefont {Kong}, \citenamefont {Zhang}, \citenamefont {Ni},
  \citenamefont {Song}, \citenamefont {Shi}, \citenamefont {Zhao},
  \citenamefont {Deng},\ and\ \citenamefont {Chen}}]{huang_2023}%
  \BibitemOpen
  \bibfield  {author} {\bibinfo {author} {\bibfnamefont {C.}~\bibnamefont
  {Huang}}, \bibinfo {author} {\bibfnamefont {T.}~\bibnamefont {Wang}},
  \bibinfo {author} {\bibfnamefont {F.}~\bibnamefont {Wu}}, \bibinfo {author}
  {\bibfnamefont {D.}~\bibnamefont {Ding}}, \bibinfo {author} {\bibfnamefont
  {Q.}~\bibnamefont {Ye}}, \bibinfo {author} {\bibfnamefont {L.}~\bibnamefont
  {Kong}}, \bibinfo {author} {\bibfnamefont {F.}~\bibnamefont {Zhang}},
  \bibinfo {author} {\bibfnamefont {X.}~\bibnamefont {Ni}}, \bibinfo {author}
  {\bibfnamefont {Z.}~\bibnamefont {Song}}, \bibinfo {author} {\bibfnamefont
  {Y.}~\bibnamefont {Shi}}, \bibinfo {author} {\bibfnamefont {H.-H.}\
  \bibnamefont {Zhao}}, \bibinfo {author} {\bibfnamefont {C.}~\bibnamefont
  {Deng}}, \ and\ \bibinfo {author} {\bibfnamefont {J.}~\bibnamefont {Chen}},\
  }\href {\doibase 10.1103/PhysRevLett.130.070601} {\bibfield  {journal}
  {\bibinfo  {journal} {Phys. Rev. Lett.}\ }\textbf {\bibinfo {volume} {130}},\
  \bibinfo {pages} {070601} (\bibinfo {year} {2023})}\BibitemShut {NoStop}%
\bibitem [{\citenamefont {Moskalenko}\ \emph {et~al.}(2021)\citenamefont
  {Moskalenko}, \citenamefont {Besedin}, \citenamefont {Simakov},\ and\
  \citenamefont {Ustinov}}]{moska2021}%
  \BibitemOpen
  \bibfield  {author} {\bibinfo {author} {\bibfnamefont {I.~N.}\ \bibnamefont
  {Moskalenko}}, \bibinfo {author} {\bibfnamefont {I.~S.}\ \bibnamefont
  {Besedin}}, \bibinfo {author} {\bibfnamefont {I.~A.}\ \bibnamefont
  {Simakov}}, \ and\ \bibinfo {author} {\bibfnamefont {A.~V.}\ \bibnamefont
  {Ustinov}},\ }\href {\doibase 10.1063/5.0064800} {\bibfield  {journal}
  {\bibinfo  {journal} {Applied Physics Letters}\ }\textbf {\bibinfo {volume}
  {119}},\ \bibinfo {pages} {194001} (\bibinfo {year} {2021})},\ \Eprint
  {http://arxiv.org/abs/https://pubs.aip.org/aip/apl/article-pdf/doi/10.1063/5.0064800/13098797/194001\_1\_online.pdf}
  {https://pubs.aip.org/aip/apl/article-pdf/doi/10.1063/5.0064800/13098797/194001\_1\_online.pdf}
  \BibitemShut {NoStop}%
\bibitem [{\citenamefont {Moskalenko}\ \emph {et~al.}(2022)\citenamefont
  {Moskalenko}, \citenamefont {Simakov}, \citenamefont {Abramov}, \citenamefont
  {Grigorev}, \citenamefont {Moskalev}, \citenamefont {Pishchimova},
  \citenamefont {Smirnov}, \citenamefont {Zikiy}, \citenamefont {Rodionov},\
  and\ \citenamefont {Besedin}}]{moska_2022}%
  \BibitemOpen
  \bibfield  {author} {\bibinfo {author} {\bibfnamefont {I.~N.}\ \bibnamefont
  {Moskalenko}}, \bibinfo {author} {\bibfnamefont {I.~A.}\ \bibnamefont
  {Simakov}}, \bibinfo {author} {\bibfnamefont {N.~N.}\ \bibnamefont
  {Abramov}}, \bibinfo {author} {\bibfnamefont {A.~A.}\ \bibnamefont
  {Grigorev}}, \bibinfo {author} {\bibfnamefont {D.~O.}\ \bibnamefont
  {Moskalev}}, \bibinfo {author} {\bibfnamefont {A.~A.}\ \bibnamefont
  {Pishchimova}}, \bibinfo {author} {\bibfnamefont {N.~S.}\ \bibnamefont
  {Smirnov}}, \bibinfo {author} {\bibfnamefont {E.~V.}\ \bibnamefont {Zikiy}},
  \bibinfo {author} {\bibfnamefont {I.~A.}\ \bibnamefont {Rodionov}}, \ and\
  \bibinfo {author} {\bibfnamefont {I.~S.}\ \bibnamefont {Besedin}},\ }\href
  {\doibase 10.1038/s41534-022-00644-x} {\bibfield  {journal} {\bibinfo
  {journal} {npj Quantum Information}\ }\textbf {\bibinfo {volume} {8}},\
  \bibinfo {pages} {130} (\bibinfo {year} {2022})}\BibitemShut {NoStop}%
\bibitem [{\citenamefont {Poletto}\ \emph {et~al.}(2012)\citenamefont
  {Poletto}, \citenamefont {Gambetta}, \citenamefont {Merkel}, \citenamefont
  {Smolin}, \citenamefont {Chow}, \citenamefont {C\'orcoles}, \citenamefont
  {Keefe}, \citenamefont {Rothwell}, \citenamefont {Rozen}, \citenamefont
  {Abraham}, \citenamefont {Rigetti},\ and\ \citenamefont
  {Steffen}}]{poletto_2012}%
  \BibitemOpen
  \bibfield  {author} {\bibinfo {author} {\bibfnamefont {S.}~\bibnamefont
  {Poletto}}, \bibinfo {author} {\bibfnamefont {J.~M.}\ \bibnamefont
  {Gambetta}}, \bibinfo {author} {\bibfnamefont {S.~T.}\ \bibnamefont
  {Merkel}}, \bibinfo {author} {\bibfnamefont {J.~A.}\ \bibnamefont {Smolin}},
  \bibinfo {author} {\bibfnamefont {J.~M.}\ \bibnamefont {Chow}}, \bibinfo
  {author} {\bibfnamefont {A.~D.}\ \bibnamefont {C\'orcoles}}, \bibinfo
  {author} {\bibfnamefont {G.~A.}\ \bibnamefont {Keefe}}, \bibinfo {author}
  {\bibfnamefont {M.~B.}\ \bibnamefont {Rothwell}}, \bibinfo {author}
  {\bibfnamefont {J.~R.}\ \bibnamefont {Rozen}}, \bibinfo {author}
  {\bibfnamefont {D.~W.}\ \bibnamefont {Abraham}}, \bibinfo {author}
  {\bibfnamefont {C.}~\bibnamefont {Rigetti}}, \ and\ \bibinfo {author}
  {\bibfnamefont {M.}~\bibnamefont {Steffen}},\ }\href {\doibase
  10.1103/PhysRevLett.109.240505} {\bibfield  {journal} {\bibinfo  {journal}
  {Phys. Rev. Lett.}\ }\textbf {\bibinfo {volume} {109}},\ \bibinfo {pages}
  {240505} (\bibinfo {year} {2012})}\BibitemShut {NoStop}%
\bibitem [{\citenamefont {Roth}\ \emph {et~al.}(2017)\citenamefont {Roth},
  \citenamefont {Ganzhorn}, \citenamefont {Moll}, \citenamefont {Filipp},
  \citenamefont {Salis},\ and\ \citenamefont {Schmidt}}]{roth2017}%
  \BibitemOpen
  \bibfield  {author} {\bibinfo {author} {\bibfnamefont {M.}~\bibnamefont
  {Roth}}, \bibinfo {author} {\bibfnamefont {M.}~\bibnamefont {Ganzhorn}},
  \bibinfo {author} {\bibfnamefont {N.}~\bibnamefont {Moll}}, \bibinfo {author}
  {\bibfnamefont {S.}~\bibnamefont {Filipp}}, \bibinfo {author} {\bibfnamefont
  {G.}~\bibnamefont {Salis}}, \ and\ \bibinfo {author} {\bibfnamefont
  {S.}~\bibnamefont {Schmidt}},\ }\href {\doibase 10.1103/PhysRevA.96.062323}
  {\bibfield  {journal} {\bibinfo  {journal} {Phys. Rev. A}\ }\textbf {\bibinfo
  {volume} {96}},\ \bibinfo {pages} {062323} (\bibinfo {year}
  {2017})}\BibitemShut {NoStop}%
\bibitem [{\citenamefont {Nesterov}\ \emph {et~al.}(2021)\citenamefont
  {Nesterov}, \citenamefont {Ficheux}, \citenamefont {Manucharyan},\ and\
  \citenamefont {Vavilov}}]{nesterov2021}%
  \BibitemOpen
  \bibfield  {author} {\bibinfo {author} {\bibfnamefont {K.~N.}\ \bibnamefont
  {Nesterov}}, \bibinfo {author} {\bibfnamefont {Q.}~\bibnamefont {Ficheux}},
  \bibinfo {author} {\bibfnamefont {V.~E.}\ \bibnamefont {Manucharyan}}, \ and\
  \bibinfo {author} {\bibfnamefont {M.~G.}\ \bibnamefont {Vavilov}},\ }\href
  {\doibase 10.1103/PRXQuantum.2.020345} {\bibfield  {journal} {\bibinfo
  {journal} {PRX Quantum}\ }\textbf {\bibinfo {volume} {2}},\ \bibinfo {pages}
  {020345} (\bibinfo {year} {2021})}\BibitemShut {NoStop}%
\bibitem [{\citenamefont {Kohler}\ \emph {et~al.}(1998)\citenamefont {Kohler},
  \citenamefont {Utermann}, \citenamefont {H\"anggi},\ and\ \citenamefont
  {Dittrich}}]{kohler_1998}%
  \BibitemOpen
  \bibfield  {author} {\bibinfo {author} {\bibfnamefont {S.}~\bibnamefont
  {Kohler}}, \bibinfo {author} {\bibfnamefont {R.}~\bibnamefont {Utermann}},
  \bibinfo {author} {\bibfnamefont {P.}~\bibnamefont {H\"anggi}}, \ and\
  \bibinfo {author} {\bibfnamefont {T.}~\bibnamefont {Dittrich}},\ }\href
  {\doibase 10.1103/PhysRevE.58.7219} {\bibfield  {journal} {\bibinfo
  {journal} {Phys. Rev. E}\ }\textbf {\bibinfo {volume} {58}},\ \bibinfo
  {pages} {7219} (\bibinfo {year} {1998})}\BibitemShut {NoStop}%
\bibitem [{\citenamefont {Hausinger}\ and\ \citenamefont
  {Grifoni}(2010{\natexlab{a}})}]{hausinger_2010}%
  \BibitemOpen
  \bibfield  {author} {\bibinfo {author} {\bibfnamefont {J.}~\bibnamefont
  {Hausinger}}\ and\ \bibinfo {author} {\bibfnamefont {M.}~\bibnamefont
  {Grifoni}},\ }\href {\doibase 10.1103/PhysRevA.81.022117} {\bibfield
  {journal} {\bibinfo  {journal} {Phys. Rev. A}\ }\textbf {\bibinfo {volume}
  {81}},\ \bibinfo {pages} {022117} (\bibinfo {year}
  {2010}{\natexlab{a}})}\BibitemShut {NoStop}%
\bibitem [{\citenamefont {Yan}\ \emph {et~al.}(2013)\citenamefont {Yan},
  \citenamefont {Gustavsson}, \citenamefont {Bylander}, \citenamefont {Jin},
  \citenamefont {Yoshihara}, \citenamefont {Cory}, \citenamefont {Nakamura},
  \citenamefont {Orlando},\ and\ \citenamefont {Oliver}}]{yan_2013}%
  \BibitemOpen
  \bibfield  {author} {\bibinfo {author} {\bibfnamefont {F.}~\bibnamefont
  {Yan}}, \bibinfo {author} {\bibfnamefont {S.}~\bibnamefont {Gustavsson}},
  \bibinfo {author} {\bibfnamefont {J.}~\bibnamefont {Bylander}}, \bibinfo
  {author} {\bibfnamefont {X.}~\bibnamefont {Jin}}, \bibinfo {author}
  {\bibfnamefont {F.}~\bibnamefont {Yoshihara}}, \bibinfo {author}
  {\bibfnamefont {D.~G.}\ \bibnamefont {Cory}}, \bibinfo {author}
  {\bibfnamefont {Y.}~\bibnamefont {Nakamura}}, \bibinfo {author}
  {\bibfnamefont {T.~P.}\ \bibnamefont {Orlando}}, \ and\ \bibinfo {author}
  {\bibfnamefont {W.~D.}\ \bibnamefont {Oliver}},\ }\href
  {https://doi.org/10.1038/ncomms3337} {\bibfield  {journal} {\bibinfo
  {journal} {Nature Communications}\ }\textbf {\bibinfo {volume} {4}},\
  \bibinfo {pages} {2337 EP } (\bibinfo {year} {2013})}\BibitemShut {NoStop}%
\bibitem [{\citenamefont {Yoshihara}\ \emph {et~al.}(2014)\citenamefont
  {Yoshihara}, \citenamefont {Nakamura}, \citenamefont {Yan}, \citenamefont
  {Gustavsson}, \citenamefont {Bylander}, \citenamefont {Oliver},\ and\
  \citenamefont {Tsai}}]{yoshihara_2014}%
  \BibitemOpen
  \bibfield  {author} {\bibinfo {author} {\bibfnamefont {F.}~\bibnamefont
  {Yoshihara}}, \bibinfo {author} {\bibfnamefont {Y.}~\bibnamefont {Nakamura}},
  \bibinfo {author} {\bibfnamefont {F.}~\bibnamefont {Yan}}, \bibinfo {author}
  {\bibfnamefont {S.}~\bibnamefont {Gustavsson}}, \bibinfo {author}
  {\bibfnamefont {J.}~\bibnamefont {Bylander}}, \bibinfo {author}
  {\bibfnamefont {W.~D.}\ \bibnamefont {Oliver}}, \ and\ \bibinfo {author}
  {\bibfnamefont {J.-S.}\ \bibnamefont {Tsai}},\ }\href {\doibase
  10.1103/PhysRevB.89.020503} {\bibfield  {journal} {\bibinfo  {journal} {Phys.
  Rev. B}\ }\textbf {\bibinfo {volume} {89}},\ \bibinfo {pages} {020503}
  (\bibinfo {year} {2014})}\BibitemShut {NoStop}%
\bibitem [{\citenamefont {Grifoni}\ and\ \citenamefont
  {H{\"a}nggi}(1998)}]{grifoni_1998}%
  \BibitemOpen
  \bibfield  {author} {\bibinfo {author} {\bibfnamefont {M.}~\bibnamefont
  {Grifoni}}\ and\ \bibinfo {author} {\bibfnamefont {P.}~\bibnamefont
  {H{\"a}nggi}},\ }\href {\doibase
  https://doi.org/10.1016/S0370-1573(98)00022-2} {\bibfield  {journal}
  {\bibinfo  {journal} {Physics Reports}\ }\textbf {\bibinfo {volume} {304}},\
  \bibinfo {pages} {229 } (\bibinfo {year} {1998})}\BibitemShut {NoStop}%
\bibitem [{\citenamefont {Hausinger}\ and\ \citenamefont
  {Grifoni}(2010{\natexlab{b}})}]{grifoni_2010}%
  \BibitemOpen
  \bibfield  {author} {\bibinfo {author} {\bibfnamefont {J.}~\bibnamefont
  {Hausinger}}\ and\ \bibinfo {author} {\bibfnamefont {M.}~\bibnamefont
  {Grifoni}},\ }\href {\doibase 10.1103/PhysRevA.81.022117} {\bibfield
  {journal} {\bibinfo  {journal} {Phys. Rev. A}\ }\textbf {\bibinfo {volume}
  {81}},\ \bibinfo {pages} {022117} (\bibinfo {year}
  {2010}{\natexlab{b}})}\BibitemShut {NoStop}%
\bibitem [{\citenamefont {Kohler}\ \emph {et~al.}(1997)\citenamefont {Kohler},
  \citenamefont {Dittrich},\ and\ \citenamefont {H\"anggi}}]{kohler_1997}%
  \BibitemOpen
  \bibfield  {author} {\bibinfo {author} {\bibfnamefont {S.}~\bibnamefont
  {Kohler}}, \bibinfo {author} {\bibfnamefont {T.}~\bibnamefont {Dittrich}}, \
  and\ \bibinfo {author} {\bibfnamefont {P.}~\bibnamefont {H\"anggi}},\ }\href
  {\doibase 10.1103/PhysRevE.55.300} {\bibfield  {journal} {\bibinfo  {journal}
  {Phys. Rev. E}\ }\textbf {\bibinfo {volume} {55}},\ \bibinfo {pages} {300}
  (\bibinfo {year} {1997})}\BibitemShut {NoStop}%
\bibitem [{\citenamefont {Breuer}\ \emph {et~al.}(2000)\citenamefont {Breuer},
  \citenamefont {Huber},\ and\ \citenamefont {Petruccione}}]{breuer_2000}%
  \BibitemOpen
  \bibfield  {author} {\bibinfo {author} {\bibfnamefont {H.-P.}\ \bibnamefont
  {Breuer}}, \bibinfo {author} {\bibfnamefont {W.}~\bibnamefont {Huber}}, \
  and\ \bibinfo {author} {\bibfnamefont {F.}~\bibnamefont {Petruccione}},\
  }\href {\doibase 10.1103/PhysRevE.61.4883} {\bibfield  {journal} {\bibinfo
  {journal} {Phys. Rev. E}\ }\textbf {\bibinfo {volume} {61}},\ \bibinfo
  {pages} {4883} (\bibinfo {year} {2000})}\BibitemShut {NoStop}%
\bibitem [{\citenamefont {Hone}\ \emph {et~al.}(2009)\citenamefont {Hone},
  \citenamefont {Ketzmerick},\ and\ \citenamefont {Kohn}}]{hone_2009}%
  \BibitemOpen
  \bibfield  {author} {\bibinfo {author} {\bibfnamefont {D.~W.}\ \bibnamefont
  {Hone}}, \bibinfo {author} {\bibfnamefont {R.}~\bibnamefont {Ketzmerick}}, \
  and\ \bibinfo {author} {\bibfnamefont {W.}~\bibnamefont {Kohn}},\ }\href
  {\doibase 10.1103/PhysRevE.79.051129} {\bibfield  {journal} {\bibinfo
  {journal} {Phys. Rev. E}\ }\textbf {\bibinfo {volume} {79}},\ \bibinfo
  {pages} {051129} (\bibinfo {year} {2009})}\BibitemShut {NoStop}%
\bibitem [{\citenamefont {Gasparinetti}\ \emph {et~al.}(2013)\citenamefont
  {Gasparinetti}, \citenamefont {Solinas}, \citenamefont {Pugnetti},
  \citenamefont {Fazio},\ and\ \citenamefont {Pekola}}]{gasparinetti_2013}%
  \BibitemOpen
  \bibfield  {author} {\bibinfo {author} {\bibfnamefont {S.}~\bibnamefont
  {Gasparinetti}}, \bibinfo {author} {\bibfnamefont {P.}~\bibnamefont
  {Solinas}}, \bibinfo {author} {\bibfnamefont {S.}~\bibnamefont {Pugnetti}},
  \bibinfo {author} {\bibfnamefont {R.}~\bibnamefont {Fazio}}, \ and\ \bibinfo
  {author} {\bibfnamefont {J.~P.}\ \bibnamefont {Pekola}},\ }\href {\doibase
  10.1103/PhysRevLett.110.150403} {\bibfield  {journal} {\bibinfo  {journal}
  {Phys. Rev. Lett.}\ }\textbf {\bibinfo {volume} {110}},\ \bibinfo {pages}
  {150403} (\bibinfo {year} {2013})}\BibitemShut {NoStop}%
\bibitem [{\citenamefont {Gasparinetti}\ \emph {et~al.}(2014)\citenamefont
  {Gasparinetti}, \citenamefont {Solinas}, \citenamefont {Braggio},\ and\
  \citenamefont {Sassetti}}]{gasparinetti_2014}%
  \BibitemOpen
  \bibfield  {author} {\bibinfo {author} {\bibfnamefont {S.}~\bibnamefont
  {Gasparinetti}}, \bibinfo {author} {\bibfnamefont {P.}~\bibnamefont
  {Solinas}}, \bibinfo {author} {\bibfnamefont {A.}~\bibnamefont {Braggio}}, \
  and\ \bibinfo {author} {\bibfnamefont {M.}~\bibnamefont {Sassetti}},\
  }\href@noop {} {\bibfield  {journal} {\bibinfo  {journal} {New Journal of
  Physics}\ }\textbf {\bibinfo {volume} {16}},\ \bibinfo {pages} {115001}
  (\bibinfo {year} {2014})}\BibitemShut {NoStop}%
\bibitem [{\citenamefont {Ferr\'on}\ and\ \citenamefont
  {Dom\'{\i}nguez}(2010)}]{ferron_2010}%
  \BibitemOpen
  \bibfield  {author} {\bibinfo {author} {\bibfnamefont {A.}~\bibnamefont
  {Ferr\'on}}\ and\ \bibinfo {author} {\bibfnamefont {D.}~\bibnamefont
  {Dom\'{\i}nguez}},\ }\href {\doibase 10.1103/PhysRevB.81.104505} {\bibfield
  {journal} {\bibinfo  {journal} {Phys. Rev. B}\ }\textbf {\bibinfo {volume}
  {81}},\ \bibinfo {pages} {104505} (\bibinfo {year} {2010})}\BibitemShut
  {NoStop}%
\bibitem [{com()}]{comment}%
  \BibitemOpen
  \href@noop {} {}\bibinfo {note} {The relaxation rate decreases with $A$
  within the range of application for the proposed gates, as seen in Figs. 6
  and 7. For much larger values, $A/\omega \gg 1$, it has an oscillatory
  behavior.}\BibitemShut {Stop}%
\bibitem [{\citenamefont {Gramajo}\ \emph {et~al.}(2018)\citenamefont
  {Gramajo}, \citenamefont {Dom\'{\i}nguez},\ and\ \citenamefont
  {S\'anchez}}]{gramajo_2018}%
  \BibitemOpen
  \bibfield  {author} {\bibinfo {author} {\bibfnamefont {A.~L.}\ \bibnamefont
  {Gramajo}}, \bibinfo {author} {\bibfnamefont {D.}~\bibnamefont
  {Dom\'{\i}nguez}}, \ and\ \bibinfo {author} {\bibfnamefont {M.~J.}\
  \bibnamefont {S\'anchez}},\ }\href {\doibase 10.1103/PhysRevA.98.042337}
  {\bibfield  {journal} {\bibinfo  {journal} {Phys. Rev. A}\ }\textbf {\bibinfo
  {volume} {98}},\ \bibinfo {pages} {042337} (\bibinfo {year}
  {2018})}\BibitemShut {NoStop}%
\bibitem [{\citenamefont {Magnus}(1954)}]{magnus_1954}%
  \BibitemOpen
  \bibfield  {author} {\bibinfo {author} {\bibfnamefont {W.}~\bibnamefont
  {Magnus}},\ }\href {http://cds.cern.ch/record/433520} {\bibfield  {journal}
  {\bibinfo  {journal} {Commun. Pure Appl. Math.}\ }\textbf {\bibinfo {volume}
  {7}},\ \bibinfo {pages} {649} (\bibinfo {year} {1954})}\BibitemShut {NoStop}%
\bibitem [{\citenamefont {Blanes}\ \emph {et~al.}(2009)\citenamefont {Blanes},
  \citenamefont {Casas}, \citenamefont {Oteo},\ and\ \citenamefont
  {Ros}}]{blanes_2009}%
  \BibitemOpen
  \bibfield  {author} {\bibinfo {author} {\bibfnamefont {S.}~\bibnamefont
  {Blanes}}, \bibinfo {author} {\bibfnamefont {F.}~\bibnamefont {Casas}},
  \bibinfo {author} {\bibfnamefont {J.}~\bibnamefont {Oteo}}, \ and\ \bibinfo
  {author} {\bibfnamefont {J.}~\bibnamefont {Ros}},\ }\href {\doibase
  https://doi.org/10.1016/j.physrep.2008.11.001} {\bibfield  {journal}
  {\bibinfo  {journal} {Physics Reports}\ }\textbf {\bibinfo {volume} {470}},\
  \bibinfo {pages} {151} (\bibinfo {year} {2009})}\BibitemShut {NoStop}%
\bibitem [{\citenamefont {Hatano}\ and\ \citenamefont
  {Suzuki}(2005{\natexlab{b}})}]{Suzuki_2005}%
  \BibitemOpen
  \bibfield  {author} {\bibinfo {author} {\bibfnamefont {N.}~\bibnamefont
  {Hatano}}\ and\ \bibinfo {author} {\bibfnamefont {M.}~\bibnamefont
  {Suzuki}},\ }in\ \href@noop {} {\emph {\bibinfo {booktitle} {Quantum
  annealing and other optimization methods}}}\ (\bibinfo  {publisher}
  {Springer},\ \bibinfo {year} {2005})\ pp.\ \bibinfo {pages}
  {37--68}\BibitemShut {NoStop}%
\end{thebibliography}%

\end{document}